\documentclass[a4paper,12pt]{article}
\pdfoutput=1
\pdfoutput=1
\usepackage{jheppub}
\usepackage{amsthm}
\usepackage{graphicx}
\usepackage{color}
\usepackage{amsmath}
\usepackage{graphicx,textcomp,float,gensymb,wrapfig, enumitem,comment,dsfont,subfigure,framed,slashed,appendix,wrapfig,wasysym}
\usepackage{comment}
\usepackage[export]{adjustbox}
\usepackage[utf8]{inputenc} 

\newcommand{\be}{\begin{equation}}
\newcommand{\ee}{\end{equation}}
\def\bsp#1\esp{\begin{split}#1\end{split}}

\newcommand{\bea}{\begin{eqnarray}}  
\newcommand{\eea}{\end{eqnarray}}

\title{Gravity-mediated Scalar Dark Matter in Warped Extra-Dimensions}
\author[a]{Miguel G. Folgado,}
\author[a]{Andrea Donini,}
\author[a]{Nuria Rius}
\affiliation[a]{Departamento de F\'isica Te\'orica and IFIC, Universidad de Valencia-CSIC, \\
C/ Catedr\'atico Jos\'e Beltr\'an, 2, E-46980 Paterna, Spain}

\emailAdd{migarfol@ific.uv.es,donini@ific.uv.es,nuria.rius@ific.uv.es}

\keywords{}

\abstract{We revisit the case of scalar Dark Matter interacting just gravitationally with the Standard Model (SM) particles in an extra-dimensional Randall-Sundrum  scenario.
We assume that both, the Dark Matter and the Standard Model, are localized in the TeV brane and only interact via gravitational mediators, namely the graviton Kaluza-Klein modes and 
the radion.  We analyze in detail the dark matter annihilation channel into two on-shell KK-gravitons, and contrary to previous studies which overlooked this 
process,  we find that it is possible to obtain the correct 
relic abundance for dark matter masses in the range [1, 10] TeV even after taking into account the strong bounds from LHC Run II. We also consider the impact of
the radion contribution (virtual exchange leading to SM final states  as well as on-shell production), which does not  significantly change our results. 
Quite interestingly, a sizeable part of the currently allowed parameter space could be tested by LHC Run III and by the High-Luminosity LHC.}

\begin{document}
\hfill {\tt FTUV-19-0523.7679,  IFIC/19-26}

\maketitle
\flushbottom

\section{Introduction}
\label{sec:intro}

The Nature of Dark Matter (DM) is one of the long-standing puzzles that still have to be explained in order to claim that we have a ``complete" picture of 
the Universe. On the one side, both from astrophysical and cosmological data (see, e.g., Ref.~\cite{Bertone:2004pz} and refs. therein), 
rather clear indications regarding the existence of some kind of matter that gravitates but
that does not interact with other particles by any other detectable mean can be gathered. On the other hand, no candidate to fill the r\^ole of Dark Matter has yet been observed
in high-energy experiments at colliders, nor is present in the Standard Model (SM) spectrum. Within SM particles, the only ones that share with Dark Matter 
the property of being weakly coupled to SM matter are neutrinos. However, experimental searches have shown that neutrinos constitute just a tiny fraction of what is called non-baryonic matter in the Universe energy budget \cite{Tanabashi:2018oca}.
Most of the suggestions for physics Beyond the Standard Model (BSM), therefore, include one or several possible candidates to be the Dark Matter. 
Under the assumptions of  the ``WIMP paradigm" (with ``WIMP" standing for ``weakly interacting massive particle"), 
these new particles have in common to be rather heavy and with very weak interactions with SM particles.
Two examples of these are the neutralino in supersymmetric extensions of the SM
\cite{Dimopoulos:1981zb} or the lightest Kaluza-Klein particle in Universal Extra-Dimensions \cite{Appelquist:2000nn,Arun:2017zap}. Searches for these heavy particles at the LHC have pushed
bounds on the masses of the candidates to the TeV range, a region of the parameter space rather difficult to test for experiments searching for Dark Matter
particles interacting directly within the detector (see, e.g., Ref.~\cite{Cushman:2013zza}) or looking at annihilation products of Dark Matter particles \cite{Cirelli:2010xx}. 
Both for this reason and for the fact that very heavy WIMP's are relatively unnatural in theories that want to solve the hierarchy problem and not 
only host some Dark Matter candidates, models in which the Dark Matter particles are either ``feebly interacting massive particles" (FIMP's) \cite{Hall:2009bx} 
or ``axion-like"  very light particles (see, e.g., Ref.~\cite{Dias:2014osa}) have been constructed. 
As a result, at present a very rich (and complicated) landscape of models explaining the Nature of Dark Matter exists, and experimental searches 
have to look for very different signals. 

In this paper we want to explore in some detail a possibility that was advanced in the literature several times in the last ten to twenty years. The idea
is that the interaction between Dark Matter particles and the SM ones, though only gravitational, may be enhanced due to the fact that gravity feels more than 
the standard $3+1$ space-time dimensions. Extra-dimensional models have been proposed to solve the hierarchy problem, related to the large hierarchy existing 
between the electro-weak scale, $\Lambda_{\rm EW} \sim 250$ GeV, and the Planck scale, $M_P \sim 10^{19}$ GeV. In all these models, 
the gravitational interaction strength is generically enhanced with respect to the standard picture since the ``true" scale of gravitation is not given by $M_P$ but, rather, by some fundamental scale $M_D$ (where $D$ is the number of dimensions). The two scales, $M_P$ and $M_D$ are connected by some relation that takes into account the geometry of space-time. In so-called {\em Large Extra-Dimensions} models (LED) \cite{Antoniadis:1990ew,Antoniadis:1997zg,ArkaniHamed:1998rs,Antoniadis:1998ig,ArkaniHamed:1998nn}, for example, $M_P^2 = V_d \times M_D^{2 + d}$ (where $d$ is the number of extra spatial dimensions). If the extra-dimensions are compactified in a $d$-dimensional volume $V_d$, and $V_d$ is sufficiently large, then $M_D \ll M_P$, thus solving or alleviating the hierarchy problem. In {\em warped extra-dimensions} (also called Randall-Sundrum models) \cite{Randall:1999ee,Randall:1999vf}, on the other hand, the separation between $M_P$ and $M_D$ is not very large, 
$M_P^2  = 8 \pi (M_D^3/k ) \left [ 1 - \exp (- 2 \pi r_c k) \right ]$, where $k$ is the curvature of the space-time along the extra-dimension and $r_c$ is the distance between two points in the extra-dimension. However, all physical masses have an exponential suppression with respect to $M_P$ due to the curvature $k$, $m = \exp (-2 \pi r_c k) \, m_0$. 
In this picture, $m_0$ is a fundamental mass parameter  of order $M_P$  and $m$ is the mass tested by a 4-dimensional observer. 
In the {\em ClockWork/Linear Dilaton} model (CW/LD), eventually, 
the relation between $M_P$ and $M_D$ is a combination of a volume factor, as for LED models, and a curvature factor, as for warped models \cite{Giudice:2016yja}.

The possibility that Dark Matter particles, whatever they be, have an {\em enhanced} gravitational interaction with SM particles have been studied mainly in the context
of warped extra-dimensions. The idea was first advanced in Refs.~\cite{Lee:2013bua,Lee:2014caa} and subsequently studied in 
Refs.~\cite{Han:2015cty,Rueter:2017nbk,Rizzo:2018obe,Rizzo:2018joy,Carrillo-Monteverde:2018phy}.
As already stressed, the Nature of Dark Matter is still unknown. In particular, if new particles are added to the SM spectrum to act as Dark Matter particles, their spin 
is completely undetermined. In the publications above, therefore, scalar, fermion and vector DM particles have been usually considered.
In this paper, on the other hand, we only consider scalar Dark Matter. We have been led to this decision by the fact that, 
maybe unexpectedly, we have found significant regions of the model parameter space for which 
the thermal relic abundance can be achieved and that can avoid present experimental bounds and theoretical constraints (in contrast, for example, with the conclusions
of Ref.~\cite{Rueter:2017nbk}). Interestingly enough, most of the allowed parameter space will be tested by the Run III at the LHC and by its high-luminosity version, the HL-LHC. 
On the way to achieve the correct relic abundance, we have found some discrepancies with existing literature on the subject when looking for DM annihilation 
into Kaluza-Klein gravitons. In addition, in order to give a consistent picture of this possibility in the framework of warped extra-dimensions, we have also taken into account the DM annihilation through and to radions within the Goldberger-Wise approach \cite{Goldberger:1999uk}, finding that this channel may also give the correct relic abundance, 
though in a very tiny region of the parameter space difficult to test at the LHC.

In forthcoming publications we plan to extend our study to DM particles with a different spin and  explore other extra-dimensional scenarios, such as LED and CW/LD. 

The paper is organized as follows: in Sec.\ref{sec:RS} we outline the theoretical framework, reminding shortly the basic ingredients of warped extra-dimensional 
scenarios and of how dark matter can be included within this hypothesis; in Sec.\ref{sec:annihilres} we show our results for the annihilation cross-sections
of scalar DM particles into SM particles, KK-gravitons and/or radions;  in Sec.\ref{sec:bounds} we review the present experimental bounds on the Kaluza-Klein graviton mass from LEP and LHC, as well as on the DM mass from direct and indirect search experiments, and we remind the theoretical constraints 
coming from unitarity violation and effective field theory consistency; in Sec.\ref{sec:Results} we explore the allowed parameter space such that 
the correct relic abundance is achieved for scalar DM particles; and, eventually, in Sec.\ref{sec:concl} we conclude. 
In the Appendices we give some of the mathematical expressions used in the paper: App.~\ref{app:spin2}  contains the 
KK-graviton propagator and polarization tensor; in App.~\ref{app:feynman} we provide the Feynman rules for our  model; in App.~\ref{app:decay}
we give the expressions for the decay amplitudes of the KK-graviton and of the radion; and, eventually, in App.~\ref{app:annihil} we give the formul\ae \, for the annihilation
cross-sections of dark matter particles into Standard Model particles, KK-gravitons and radions.

\section{Theoretical framework}
\label{sec:RS}

In this Section, we shortly review the Warped Extra-Dimensions scenario (also called Randall-Sundrum model \cite{Randall:1999ee}) and 
introduce our setup to include Dark Matter in the model, 
we give the relevant formul\ae \, to compute the DM relic abundance and eventually provide the DM annihilation cross-sections 
into SM particles, Kaluza-Klein gravitons and into radions.

\subsection{A short summary on Warped Extra-Dimensions}
\label{sec:warped}

The popular Randall-Sundrum scenario (from now on RS or RS1 \cite{Randall:1999ee}, to be distinguished from the scenario called RS2 \cite{Randall:1999vf})
consider a non-factorizable 5-dimensional metric in the form: 
\be
\label{eq:5dmetric}
ds^2 = e^{-2\sigma} \eta_{\mu \nu} dx^{\mu}dx^{\nu} - r_c^2 \, dy^2
\ee
where $\sigma = k r_c |y|$ and the signature of the metric is $(+,-,-,-,-)$.
In this scenario, $k$ is the curvature along the 5th-dimension and it is ${\cal O} \left ( M_P \right )$. 
The length-scale $r_c$, on the other hand, is related to the size of the extra-dimension: we only consider a slice of the space-time between two branes
located conventionally at the two fixed-points of an orbifold, $y = 0$ (the so-called UV-brane) and $y = \pi$ (the IR-brane). 
The 5-dimensional space-time is a slice of ${\cal }AdS_5$ and 
the exponential factor that multiplies the ${\cal M}_4$ Minkowski 4-dimensional space-time is called the ``warp factor". Notice that, in order to have gravity
in 4-dimensions, in general $\eta_{\mu\nu} \to g_{\mu\nu}$, with $g_{\mu\nu}$ the 4-dimensional induced metric on the brane.

The action in 5D is: 
\begin{equation}
S = S_{\rm gravity} + S_{\rm IR} + S_{\rm UV}
\end{equation}
where
\begin{equation}
\label{eq:5dgravity}
S_{\rm gravity} = \frac{16 \pi}{M_5^3} \int d^4 x \, \int_0^\pi r_c dy \sqrt{G^{(5)}} \, \left [ R^{(5)}  - 2 \Lambda_5  \right ] \, ,
\end{equation}
with $M_5$ the fundamental gravitational scale, $G_{MN}^{(5)}$ and $R^{(5)}$ the 5-dimensional metric and Ricci scalar, respectively, and $\Lambda_5$ the 5-dimensional 
cosmological constant. As usual, we consider capital latin indices $M,N$ to run over the 5 dimensions and greek indices $\mu,\nu$ only over 4 dimensions.
The Planck mass is related to the fundamental scale $M_5$ as: 
\be
\bar M_P^2 = \frac{M_5^3}{k} \,  \left (1-e^{-2k \pi r_c} \right ) \, ,
\ee
where $\bar{M}_P = M_P/\sqrt{8\pi} = 2.435 \times 10^{18} $ GeV is the reduced Planck mass.

We consider for the two brane actions  the following expressions: 
\begin{equation}
S_{\rm IR} = \int d^4 x \sqrt{- g} \, \left \{ - f_{\rm IR}^4 + {\cal L}_{\rm SM} + {\cal L}_{\rm DM}  \right \}
\end{equation}
and
\begin{equation}
S_{\rm UV} = \int d^4 x \sqrt{-g} \, \left \{ - f_{\rm UV}^4 + \dots \right \} \, ,
\end{equation}
where $f_{\rm IR}, f_{\rm UV}$ are the brane tensions for the two branes. In Randall-Sundrum scenarios, in order to achieve the metric in eq.~(\ref{eq:5dmetric}) as a classical solution of the Einstein equations, the brane-tension terms in $S_{\rm UV}$ and $S_{\rm IR}$ are chosen such as to cancel the 5-dimensional cosmological constant, 
$f_{\rm IR}^4 = - f_{\rm UV}^4 = \sqrt{- 24 M_5^3 \, \Lambda_5}$. 

Throughout this paper, we consider all the SM and DM fields localized on the IR-brane, whereas on the UV-brane we could have
any other physics that is Planck-suppressed. We assume that DM particles only interact with the SM particles gravitationally and, for simplicity, we focus on scalar DM.
More complicated DM spectra (with particles of spin higher than zero or with several 
particles) will not be studied here. Notice that, in 4-dimensions, the gravitational interactions would be enormously suppressed by powers of the Planck mass. However, in an extra-dimensional scenario, the gravitational interaction is actually enhanced: on the IR--brane, in fact, the effective gravitational coupling is 
$\Lambda = \bar M_P \exp \left ( - k \pi r_c \right )$, due to the rescaling factor $\sqrt{G^{(5)}}/\sqrt{- g^{(4)}}$.
It is easy to see that $\Lambda \ll \bar M_P$ even for moderate choices of $\sigma$. In particular, for $\sigma = k  r_c \sim 10$ the RS scenario can address the hierarchy problem. In general, we will work with $\Lambda = {\cal O} (1 \, \rm TeV)$ but not necessarily as low as to solve the hierarchy problem.
 
Expanding the 4-dimensional component of the metric at first order about its static solution, we have: 
\be
\label{eq:metricexpansion}
G^{(5)}_{\mu\nu} =  e^{-2\sigma} (\eta_{\mu \nu} + \kappa_5 h_{\mu \nu}) \, ,
\ee
with $\kappa_5 = 2 M_5^{-2/3}$. The 5-dimensional field $h_{\mu\nu}$ can be written as a Kaluza-Klein tower of 4-dimensional fields as follows:
\be
h_{\mu \nu}(x,y) = \sum h^{n}_{\mu \nu}(x) \frac{\chi^{n}(y)}{\sqrt{r_c}} \, .
\ee
The $h^n_{\mu\nu} (x)$ can be interpreted as the KK-excitations of the 4-dimensional graviton. The $\chi^{n}(y)$ factors are the wavefunctions of the KK-gravitons 
along the extra-dimension. Notice that in the 4-dimensional decomposition of 
a 5-dimensional metric, two other fields are generally present: the graviphoton, $G^{(5)}_{\mu5}$, and the graviscalar $G^{(5)}_{55}$.
It has been shown elsewhere \cite{Randall:1999ee} that graviphotons are massive due to the breaking of 5-dimensional translational invariance induced by 
 the presence of the branes. On the other hand, the graviscalar field is relevant 
to stabilize the size of the extra-dimension and it will be discussed below when introducing the {\em radion}. 

The equation of motion for the $n$-th KK-mode is given by:
\be
(\eta^{\mu \nu}\partial_\mu \partial_\nu + m_n^2) h^{n}_{\mu \nu}(x) = 0 \, ,
\ee
where $m_n$ is its mass. Using the Einstein equations we obtain \cite{Davoudiasl:1999jd}:
\be
\frac{-1}{r_c^2} \frac{d}{dy} \left( e^{-4\sigma} \frac{d\chi^{n}}{dy} \right) = m_n^2 e^{-2\sigma}\chi^{n} \, .
\ee
from which:
\be
\chi^{n}(y)=\frac{e^{2\sigma(y)}}{N_n}\left[ J_2(z_n) + \alpha_n Y_2(z_n)\right] \, ,
\ee
being $J_2$ and $Y_2$ Bessel functions of order 2 and $z_n (y) = m_n/k e^{\sigma(y)}$.
The $N_n$ factor is the $n$-th KK-mode wavefunction normalization.
In the limit $m_n / k \ll 1$ and $e^{k \pi r_c} \gg 1$, the coefficient $\alpha_n$ becomes $\alpha_n \sim x_n^2 \exp \left (-2 k \pi r_c \right )$, 
where $x_n$ are the are the roots of the Bessel function $J_1$, $J_1 (x_n) = 0$, and the masses of the KK-graviton modes are given by: 
\be
m_n = k x_n e^{- k \pi r_c} \, .
\ee
Notice that, for low $n$, the KK-graviton masses are not equally spaced, as they are proportional to the roots
of the Bessel function $J_1$. This is very different from both the LED and the CW/LD scenarios, 
however for large $n$ the spacing between KK-graviton modes become so small 
that all extra-dimensional scenarios eventually coincide, $m_n \sim n/R$ (being $R$ some relevant length scale specific to each scenario).

The normalization factors can be computed imposing that: 
\be
\int dy e^{-2\sigma} \left [\chi^{n} \right]^2 = 1 \, .
\ee
In the same approximation as above, i.e. for $m_n / k \ll 1$ and $e^{k \pi r_c} \gg 1$, we get: 
\begin{equation}
N_0 = - \frac{1}{\sqrt{k r_c}} \qquad ; \qquad N_n = \frac{1}{\sqrt{2 k r_c}}  \, e^{k \pi r_c} \, J_2 (x_n) \, .
\end{equation}
Notice the difference between the $n=0$ mode and the $n>0$ modes:  for $n=0$, the wave-function at the IR-brane location $y=\pi$ takes the form
\be
\chi^{0}(y=\pi) = \sqrt{k r_c} \left (1-e^{-2 k \pi r_c} \right ) = -\sqrt{r_c} \, \frac{M_5^{3/2}}{\bar M_P} \, ,
\ee
whereas for $n>0$:
\be
\chi^{n}(y=\pi) = \sqrt{k r_c} \, e^{k \pi r_c} = \sqrt{r_c} \, e^{k \pi r_c} \frac{M_5^{3/2}}{\bar M_P} = \sqrt{r_c} \, \frac{M_5^{3/2}}{\Lambda}
\ee

The important difference can be easily understood by looking at the coupling between the energy-momentum tensor 
and gravity at the location of the IR-brane:
\be
\mathcal{L} = -\frac{1}{M_5^{3/2}} T^{\mu \nu}(x) h_{\mu \nu}(x,y=\pi) = -\frac{1}{M_5^{3/2}} T^{\mu \nu}(x) \sum_{n=0} h^{n}_{\mu \nu}\frac{\chi^{n}}{\sqrt{r_c}} \, ,
\ee
where the only scale is the fundamental gravitational scale $M_5$. However, if we separate the $n=0$ and the $n>0$ modes
we get:
\be
\mathcal{L} = -\frac{1}{\bar M_P} T^{\mu \nu}(x) h^{0}_{\mu \nu}(x) -\frac{1}{\Lambda} \sum_{n=1} T^{\mu \nu}(x) h^{n}_{\mu \nu}(x) \, ,
\ee
from which is clear that the coupling between KK-graviton modes with $n \neq 0$ is suppressed by the effective scale $\Lambda$
and not by the Planck scale. 

It is useful to remind here the explicit form of the energy-momentum tensor:
 \be
T_{\mu \nu} = T_{\mu \nu}^{SM} + T_{\mu \nu}^{DM} \, ,
\label{Tensor}
\ee
where
 \begin{align}
T_{\mu \nu}^{SM} =& \left[ \frac{i}{4}\bar{\psi} (\gamma_\mu D_\nu + \gamma_\nu D_\mu)\psi - \frac{i}{4}(\gamma_\mu D_\nu \bar{\psi} \gamma_\mu + D_\mu\bar{\psi}\gamma_\nu)\psi - \eta_{\mu\nu}(\bar{\psi}\gamma^\mu D_\mu \psi - m_\psi \bar{\psi}\psi) + \right. \nonumber \\
& \left. + \frac{i}{2}\eta_{\mu \nu} \partial^\rho \bar{\psi}\gamma_\rho\psi  \right] + \left[ \frac{1}{4}\eta_{\mu\nu} F^{\lambda \rho}F_{\lambda \rho} -  F_{\mu \lambda}F^{\lambda}_{\ \nu}  \right] + \left[ \eta_{\mu\nu} D^\rho H^\dagger D_\rho H + \eta_{\mu\nu} V(H) + \right. \nonumber \\ 
&\left. + D_\mu H^\dagger D_\nu H + D_\nu H^\dagger D_\mu H  \right] \nonumber \\
\hphantom{} \nonumber
\end{align}
and
\begin{equation}
T_{\mu \nu}^{DM} = \ (\partial_\mu S) (\partial_\nu S) - \frac{1}{2}\eta_{\mu \nu} (\partial^\rho S) (\partial_\rho S) + \frac{1}{2} \eta_{\mu \nu} m_S^2 S^2 \, ,
\label{Tensor_SM_DM}
\end{equation}
where we have introduced the scalar singlet field $S$ to represent the DM particle in our scenario.

Notice that a scalar DM particle will also interact with the SM  through the so-called "Higgs portal", namely 
 \begin{equation}
\label{eq:Higgsportal}
 {\cal L}_{\rm DM} \supset \lambda_{hS} (H^\dagger H) (S^\dagger S) \ , 
\end{equation}
since this term is always allowed.
However,  such coupling is strongly constrained (see Sect.~\ref{sec:relic}), and we neglect its effect in our analysis.

 \subsection{Adding the radion}
\label{sec:rad}

Stabilizing the size of the extra-dimension to be $y = \pi r_c$ is a complicated task. In general (see, e.g., Refs.~\cite{Appelquist:1982zs,Appelquist:1983vs,deWit:1988ct}) bosonic quantum loops have a net effect on the border of the extra-dimension such that the extra-dimension
itself should shrink to a point. This feature, in a flat extra-dimension, can only be compensated by fermionic quantum loops
and, usually, some supersymmetric framework is invoked to stabilize the radius of the extra-dimension (see, e.g., Ref.~\cite{Ponton:2001hq}). 
In Randall-Sundrum scenarios, on the other hand, a new mechanism has been considered: if we add a bulk scalar field $\Phi$
with a scalar potential $V(\Phi)$ and some {\rm ad hoc}  localized potential terms, $\delta (y=0) V_{\rm UV}(\Phi)$ and 
$\delta (y = \pi r_c) V_{\rm IR} (\Phi)$, it is possible to generate an effective potential $V(\varphi)$ for the four-dimensional field $\varphi = f \, \exp \left ( - k \pi T \right )$
(with $f = \sqrt{24 M_5^3/k}$ and $\langle T \rangle  = r_c$). The minimum of this potential can yield the desired value of $k r_c$ without extreme fine-tuning of the 
parameters \cite{Goldberger:1999wh,Goldberger:1999uk}.

As in the spectrum of the theory there is already a scalar field, the graviscalar $G^{(5)}_{55}$, the $\Phi$ field will generically
mix with it. The KK-tower of the graviscalar is absent from the low-energy spectrum, as they are eaten by the KK-tower 
of graviphotons to get a mass (due to the spontaneous breaking of translational invariance caused by the presence of one or more
branes). On the other hand, the KK-tower of the field $\Phi$ is present, but heavy (see Ref.~\cite{Goldberger:1999un}). 
The only light field present in the spectrum is a combination of the graviscalar zero-mode and the $\Phi$ zero-mode.
This field is usually called the {\em radion}, $r$. Its mass can be obtained from the effective potential $V (\varphi)$ and is given by
$m_\varphi^2 = k^2 v_v^2 / 3 M_5^3 \, \epsilon^2 \, \exp (-2 \pi k r_c)$, where $v_v$ is the value of $\Phi$ at the visible brane and $\epsilon = m^2/4 k^2$ (with $m$ 
the mass of the field $\Phi$). Quite generally $\epsilon \ll 1$ and, therefore, the mass of the radion can be much smaller than the first KK-graviton mass.

The radion, as for the KK-graviton case, interacts with both the DM and SM particles. It couples with matter
through the trace of the energy-momentum tensor $T$ \cite{Lee:2013bua}. Massless gauge fields do not contribute to the trace 
of the energy-momentum tensor,  but effective couplings are generated from two different sources: quarks and $W$ boson loops  and the trace anomaly 
\cite{Blum:2014jca}. Thus the radion  Lagrangian takes the following form \cite{Goldberger:1999un, Csaki:1999mp}:
\be
\mathcal{L}_r = \frac{1}{2}(\partial_\mu r)(\partial^\mu r) - \frac{1}{2} m_r^2 r^2 +  \frac{1}{\sqrt{6}\Lambda} r T+ \frac{\alpha_{EM} \, C_{EM}}{8\pi\sqrt{6}\Lambda} r F_{\mu\nu} F^{\mu\nu} + \frac{\alpha_{S}C_{3}}{8 \pi \sqrt{6} \Lambda} r \sum_a F^a_{\mu\nu} F^{a\mu\nu}\, ,
\label{radion_lag}
\ee
where $F_{\mu\nu}, F^a_{\mu\nu}$ are the Maxwell and $SU_c(3)$ Yang-Mills tensors, respectively.
The $C_3$ and $C_{EM}$ constants encode all information about the massless gauge boson contributions and are given in App. \ref{app:feynman}.

\subsection{The DM Relic Abundance in the Freeze-Out scenario}
\label{sec:relic}

Experimental data ranging from astrophysical to cosmological scales point out that a significant fraction of the Universe
energy appears in the form of a non-baryonic ({\em i.e.} electromagnetically inert) matter. This component of the 
Universe energy density is called {\em Dark Matter} and, in the cosmological ``standard model", the $\Lambda$CDM, 
it is usually assumed to be represented by stable (or long-lived) heavy particles ({\em i.e.} non-relativistic, or ``cold"). 
Within the thermal freeze-out scenario
 the DM component is supposed to be in thermal equilibrium with the rest of particles in the Early Universe. 
The evolution of the dark matter number density $n_{\rm DM}$ in this paradigm is 
governed by the Boltzmann equation \cite{Kolb:1990vq}:
\be
\frac{dn_{\rm DM}}{dt} = -3 H(T) \, n_{\rm DM} - \left\langle \sigma v \right\rangle \left [ n_{\rm DM}^2 - (n_{\rm DM}^{eq})^2 \right] \, , 
\label{boltzmann_equation}
\ee
where $T$ is the temperature,  $H(T)$ is the Hubble parameter as a function of the temperature, 
and $n_{\rm DM}^{eq}$ is the DM number density at equilibrium (see, {\em e.g.}, Ref.~\cite{Kolb:1990vq}). 
The Boltzmann equation is governed by two factors: one proportional to $H(T)$ and the second to the thermally-averaged cross-section, 
$\left\langle \sigma v \right\rangle$.  In order for $n_{\rm DM} (T)$ to freeze-out, as the Universe expanded and cooled down 
the thermally-averaged annihilation cross-section  $\left\langle \sigma v \right\rangle$ times the number density should fall below $H(T)$. 
At that moment, DM decouples from the rest of particles leaving an approximately constant number density in the co-moving frame, called relic abundance.

The experimental value of the relic abundance can be computed starting from the DM density in the $\Lambda$CDM model. 
From Ref.~\cite{Aghanim:2018eyx} we have $\Omega_{\rm CDM} h^2 = 0.1198 \pm 0.0012$, where $h$ parametrizes the present Hubble parameter. 
Solving eq.~(\ref{boltzmann_equation}), it can be found  the thermally-averaged cross-section at  freeze-out 
$\left\langle \sigma_{\rm FO} \, v \right\rangle = 2.2 \times 10^{-26}$ cm$^3$/s \cite{Steigman:2012nb}.
Notice that for $m_{\rm DM} >  10$ GeV,  the relic abundance is insensitive to the value of $m_{\rm DM}$ and therefore 
 the thermally-averaged annihilation cross section $\sigma_{\rm FO}$ needed to obtain the correct relic abundance 
is not a function of the DM particle mass. 

When comparing the prediction of a given model to the expectation in the freeze-out scenario, 
the key parameter to compute the relic abundance is, thus,  $\left\langle \sigma v \right\rangle$. 
In order to obtain this quantity, we must first calculate the total annihilation cross-section of the DM particles (represented in our case by the field $S$):
\be
\label{eq:sigmaDMtot}
\sigma_{\rm th}= \sum_{\rm SM} \sigma_{\rm ve}(S \, S  \rightarrow {\rm SM} \, {\rm SM}) 
+ \sum_{n=1} \sum_{m=1} \sigma_{GG} (S \, S \rightarrow G_{n} \, G_{m}) 
+ \sigma_{rr} (S \, S \rightarrow r \, r) \, ,
\ee
where in the first term, $\sigma_{\rm ve}$ ("${\rm ve}$" stands for "virtual exchange"), we sum over all SM particles. 
The second term, $\sigma_{GG}$, corresponds to DM annihilation into a pair of KK-gravitons, $G_n \, G_m$. 
Eventually, the third term, $\sigma_{rr}$, corresponds to DM annihilation into radions.

If the DM mass $m_S$ is smaller than the mass of the first KK-graviton and of the radion, only the first channel exists. 
Since in the freeze-out paradigm the DM particles have small relative velocity $v$ when the freeze-out occurs, 
it is useful to approximate the c.o.m. energy $s$ as $s \sim 4 m_S^2$, and keep only the leading order in the so-called {\em velocity expansion}.
Formul\ae \, for the DM annihilation into SM particles within this approximation are given in App.~\ref{app:annihil}. 

Notice that DM annihilation to SM particles can occur through three possible mediators: the Higgs boson, the KK-tower of gravitons and the radion.
The first option, that depends on the coupling introduced in eq.~(\ref{eq:Higgsportal}), has been extensively studied. Current bounds (see for instance \cite{Escudero:2016gzx,Casas:2017jjg} for recent analyses) rule out DM masses $m_S \lesssim $ 500 GeV (except for the Higgs-funnel region, $m_S \simeq m_h/2$) and future direct detection experiments such as LZ \cite{Akerib:2015cja} will either find DM or exclude larger masses, up to $\cal{O}$(TeV). In the presence of other annihilation channels, as in our case, if LZ does not get any positive signal of DM it will lead to a stringent limit on the Higgs portal coupling 
$\lambda_{hS}$, so that the Higgs boson contribution to DM annihilation into SM particles will be negligible for DM masses at the TeV scale \cite{Escudero:2016ksa, Casas:2017jjg}. In the rest of the paper, we will assume that $\lambda_{hS}$ is small enough so as to 
be irrelevant in our analysis, and we will not consider this channel any further.

On the other hand, depending on the particular values  for the radion mass (determined by the specific
features of the bulk and localized scalar potentials) and the KK-graviton masses 
(fixed by  $k, M_5$ and $r_c$), radion or KK-graviton exchange can dominate the annihilation amplitude. 
When computing the contribution of the radion and KK-graviton exchange to the DM annihilation  cross-section into SM particles, 
it is of the uttermost importance to take into account properly the decay width of the radion and of the KK-gravitons, 
respectively \footnote{
In the case of the KK-gravitons, due to the breaking of translational invariance in the extra-dimension, the KK-number is not conserved
and heavy KK-graviton modes can also decay into lighter KK-gravitons when kinematically allowed. Formul\ae \, for the radion and KK-graviton 
decays are given in App.~\ref{app:decay}.}.
Notice that  the DM annihilation cross-section into SM particles via virtual exchange of KK gravitons is velocity suppressed ($d$ wave), 
due to the spin 2 of the mediators, while the corresponding one through virtual radion is $s$ wave.

Within the Goldberger-Wise stabilization mechanism, the radion is expected to be lighter than the first KK-graviton mode, so 
the next  channel to open is usually the DM annihilation into radions. 
The analytic expression for $\sigma_{rr} (S \, S \rightarrow r \, r)$ in the approximation $s \sim 4 m_S^2$ is 
given in App.~\ref{app:annihil}.  It is also $s$ wave.

Eventually, for DM masses larger than the mass of the first KK-graviton mode, annihilation of DM particles into KK-gravitons becomes
possible and the last channel in eq.~(\ref{eq:sigmaDMtot}) opens. As the KK-number is not conserved due to the presence of the branes
in the extra-dimension (that breaks explicitly momentum conservation in the 5th-dimension), any  combination of KK-graviton modes 
is possible when kinematically allowed. Therefore, we must sum over all the modes as long as the condition $2 m_S \geq m_{G_n} + m_{G_m}$ is fulfilled. 
The analytic expression for $\sigma_{GG} (S \, S \rightarrow G_n \, G_m)$ at leading order in the velocity expansion is 
also given in App.~\ref{app:annihil}, and it turns out to be $s$ wave as well.
 Notice that we will not take into account annihilation into zero-modes gravitons, $G_0 \, G_0$ or $G_0 \, G_n$, 
as these channels are Planck-suppressed with respect to the production of a pair of massive KK-graviton modes, $G_n \, G_m$.

As the {\em velocity expansion} approximation may fail in the neighbourghood of resonances and, in the RS model, the virtual graviton exchange cross-section is indeed the result of an infinite sum of KK-graviton modes, we computed the analytical value of $\left\langle \sigma v \right\rangle$ using the exact expression from Ref.~\cite{Gondolo:1990dk}:
\be
\left\langle \sigma v \right\rangle =\frac{1}{8m_S^4TK_2^2(x)} \int_{4m_S^2}^{\infty} ds (s-4m_S^2)\, \sqrt{s} \, \sigma(s) \, K_1\left(\frac{\sqrt{s}}{T}\right)\, ,
\label{thermal_average}
\ee
where $K_1$ and $K_2$ are the modified Bessel functions and $ v$ is the relative (M\o ller) velocity of the DM particles.





\section{Scalar DM annihilation cross-section in RS}
\label{sec:annihilres}

For relatively low DM mass the only open annihilation channel  is  into SM particles through KK-graviton or radion exchange. 
Direct production of radions or KK-gravitons in the final state becomes allowed for  DM mass $m_{S} \geq m_{G_1}, m_r$, where $m_{S}$ and $m_{G_1}$ are the DM and the first KK-graviton masses, respectively.  

\begin{figure}[htbp]
\centering
\includegraphics[width=120mm]{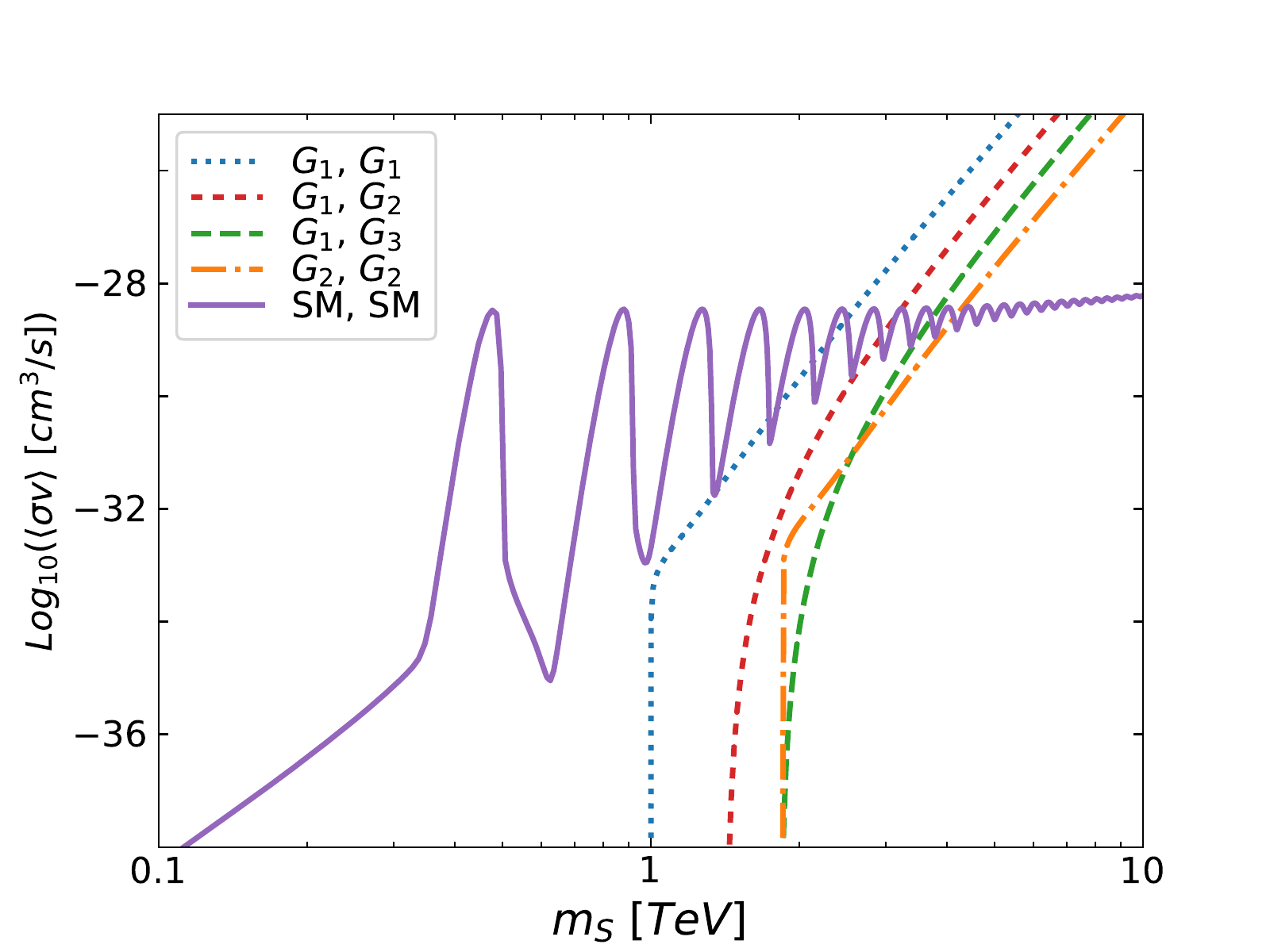}
\caption{\it 
Contributions to the scalar DM annihilation cross-section due to KK-gravitons, for $\Lambda = 100$  TeV and $m_{G_1} = 1$ TeV, as a function of the DM mass $m_{S}$.
The solid purple line corresponds to the DM annihilation into SM particles through virtual KK-graviton exchange, $\sigma_{{\rm ve},G}$. 
The non-solid lines correspond to DM annihilation into two KK-gravitons, $\sigma_{G G}$: from left to right $SS \to  (G_1, G_1), (G_1, G_2), (G_2, G_2)$ and $(G_1, G_3)$, respectively. 
}
 \label{fig:contribuciones}
\end{figure}

\subsection{Virtual KK-graviton exchange and on-shell KK-graviton production}
\label{sec:KKgravxsection}

We plot in Fig.~\ref{fig:contribuciones} 
the different KK-graviton contributions to $\left\langle \sigma v \right\rangle$ separately, so as to understand clearly the main features.

We consider first the case of DM annihilation into SM particles through KK-graviton exchange,  summed over all virtual KK-gravitons, $\sigma_{{\rm ve},G}$.
This result is shown by the solid (purple) line in Fig.~\ref{fig:contribuciones} as a function of the DM mass $m_{S}$, for the particular choice $\Lambda = 100$ TeV and $m_{G_1} = 1$ TeV. When the DM particle mass is nearly half of one of the KK-graviton masses, 
$s = 4 m_{S}^2 \sim m^2_{G_n}$, the resonant contribution dominates the cross-section, which abruptly increases.
At each of the resonances, 
$\left\langle \sigma v \right\rangle$ depends only marginally on the DM mass $m_{S}$ and, therefore, we have an approximately constant thermally-averaged maximal cross-section
(a small $m_S$-dependence  arises only at very large values of $m_S$). This contribution was studied in detail in Ref.~\cite{Rueter:2017nbk}, where it was shown that the resonant enhancement of the cross-section for $m_{S } \sim m_{G_n}/2$ was not enough to achieve the value of $\left\langle \sigma_{\rm FO} v \right\rangle$ that gives the correct relic abundance, once values of $\Lambda$ compatible with LHC exclusion bounds were taken into account.

On the other hand, for $m_{S} \geq m_{G_1}$ DM annihilation into on-shell KK-gravitons becomes possible. 
Depending on the DM particle mass, production of several KK-graviton modes is allowed. This is represented in Fig.~\ref{fig:contribuciones} 
by dashed or dot-dashed lines, where we show the contribution to the DM annihilation cross-section from the channels $SS \to (G_1 \, G_1), (G_1 \, G_2), (G_2 \, G_2)$ 
and $(G_1 \, G_3)$. More channels open for larger values of $m_{S}$ that however have not been depicted in Fig.~\ref{fig:contribuciones}, where we have decided to show just the lowest-lying ones for the sake of clarity of the plot. 
Recall that each of the two KK-gravitons can have any KK-number: in particular, it is not forbidden by any symmetry to have $SS \to G_n \, G_m$ with $n \neq m$, as translational invariance in the 5th-dimension is explicitly broken due to the presence of the IR- and UV-branes and the KK-number is not conserved. As it can be seen in the Figure,
the contribution of each channel to the total cross-section varies with the DM mass. For example, $SS \to G_2 \, G_2$ (orange, dot-dashed line)  dominates over $SS \to G_1 \, G_3$ (green, dashed line) in a very small range of $m_{S}$, whereas the latter takes over for large $m_{S}$. Notice that, although KK-graviton production was considered in Ref.~\cite{Lee:2013bua}, the possibility of producing different KK-graviton modes was overlooked there. 

\begin{figure}[htbp]
\centering
\begin{tabular}{cc}
\includegraphics[width=72mm]{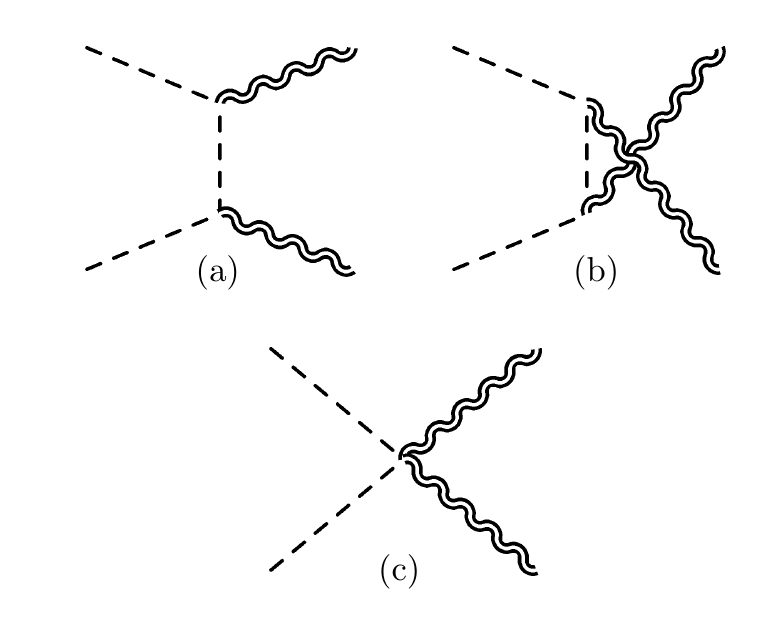} &
\includegraphics[width=72mm]{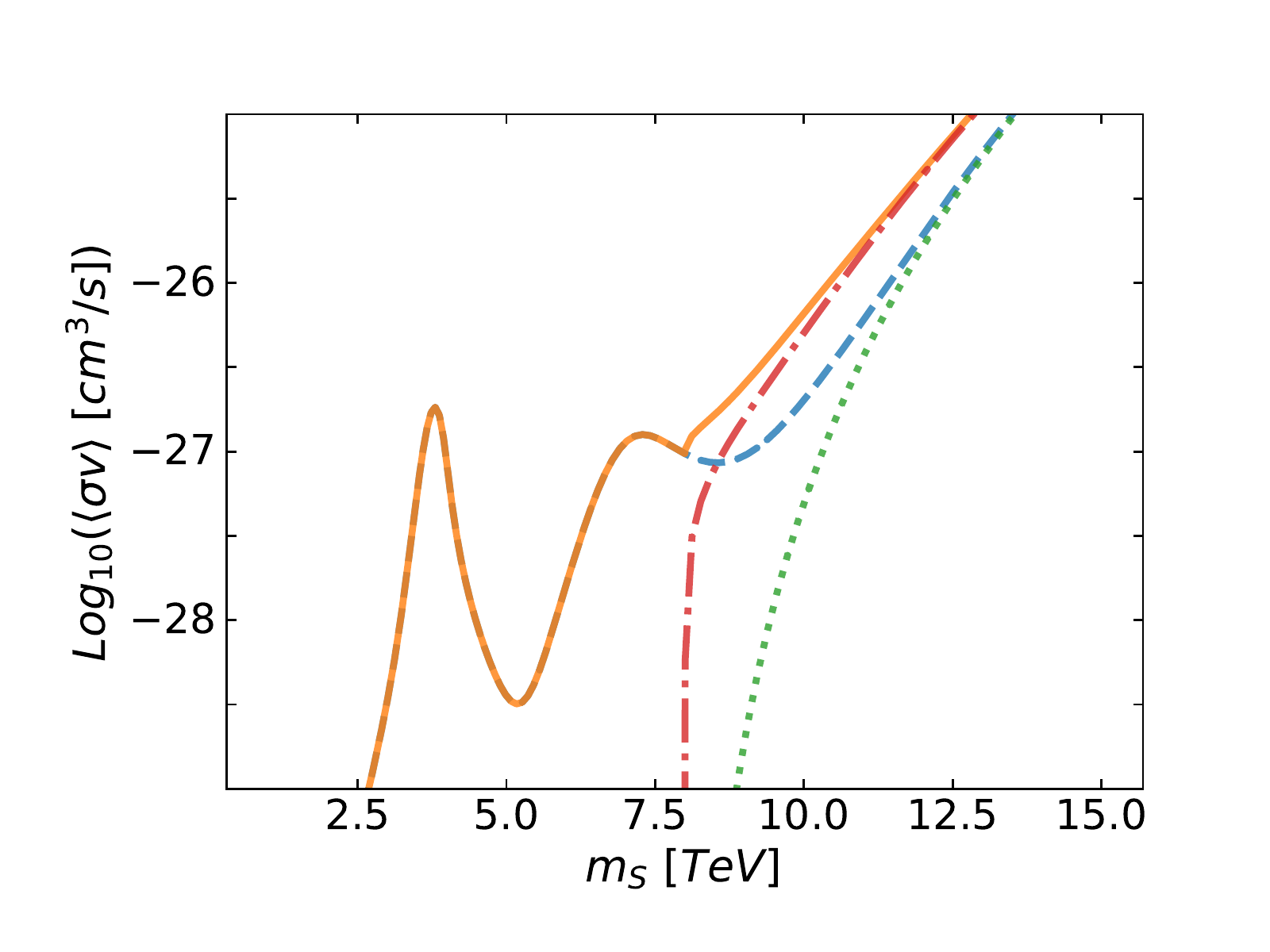}
\end{tabular}
\caption{\it 
Left panel: Feynman diagrams corresponding to the different amplitudes that contribute
to scalar DM annihilation into two on-shell KK-gravitons at ${\cal O}(1/\Lambda^2)$.
Diagrams (a) and (b): $t$- and $u$-channel DM exchange. Diagram (c): second order expansion of the metric in eq.~(\ref{eq:metricexpansion}).
Right panel: Relevance of overlooked contributions to the scalar DM annihilation cross-section for $\Lambda = 10$ TeV 
and $m_{G_1} = 8$ TeV, as a function of the DM mass $m_S$.
The solid orange (blue dashed) line corresponds to the DM annihilation cross-section through and into KK-gravitons with (without) the contribution to the amplitude
from diagram (c). The dot-dashed red (dotted green) line is the DM annihilation cross-section 
into KK-gravitons, only, with (without) the contribution from diagram (c).
}
 \label{fig:diagramas}
\end{figure}

In Fig.~\ref{fig:diagramas} (left panel) we plot the different Feynman diagrams that contribute to DM annihilation into on-shell KK-gravitons.
Diagram (c) was not considered previously in the literature (see, {\em e.g.}, Ref.~\cite{Lee:2013bua}). However, it  must be taken into account when computing the  production of two real gravitons, as the corresponding amplitude is also proportional to $1/\Lambda^2$, the  same order as  the two other 
diagrams\footnote{Notice that on-shell KK-graviton production through KK-graviton exchange
in $s$-channel only appears when expanding the metric in eq.~(\ref{eq:metricexpansion}) up to 
third order and, therefore, the corresponding amplitude is suppressed by $1/\Lambda^4$.
}. 
The corresponding Feynman rule can be obtained by expanding the metric up to second order about the Minkowski space-time:
\be
\mathcal{L} \supset - \frac{1}{2 \Lambda^2}\sum_{n=1} T^{\mu \nu}(x) \left( h^{(n)}_{\mu \alpha}(x)h^{(n)}_{\beta \nu}(x)\eta^{\alpha \beta} + h^{(n)}_{\mu \nu}(x)h^{(n)}_{\alpha\beta}(x)\eta^{\alpha \beta} \right) \, .
\label{lagrangiano_4_puntas}
\ee
Notice that, if a diagram that should be considered at a given order in $1/\Lambda$ when computing a given process is absent, then the gravitational gauge-invariance
of the amplitude is not guaranteed and the cross-section computation is built over slippery ground from a theoretical point of view.
The impact of diagram (c) is shown in Fig.~\ref{fig:diagramas} (right panel), where we compare the total DM annihilation cross-section 
through and into KK-gravitons including or not the contribution to the amplitude from this diagram, for a particular choice of $m_{G_1} = 8$ TeV and $\Lambda = 10$ TeV. 
The solid orange (blue dashed) line is the total DM annihilation cross-section through and into KK-gravitons with (without) diagram (c), whereas
the dot-dashed red (dotted green) line is the DM annihilation cross-section into KK-gravitons with (without) diagram (c). It can be seen that, for this particular 
choice of $m_{G_1}$ and $\Lambda$, the difference between the two computations can be as large a one order of magnitude for $m_S \sim 10$ TeV.

\begin{figure}[htbp]
\centering
\includegraphics[width=160mm]{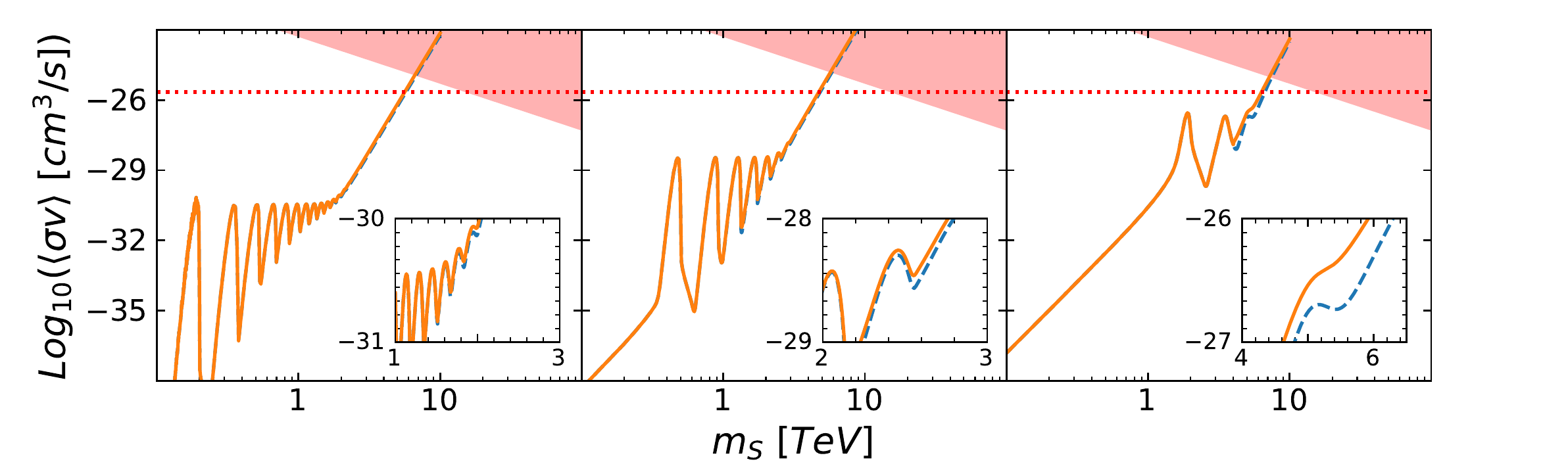}
\caption{\it The thermally-averaged scalar DM annihilation cross-section through virtual
KK-graviton exchange and direct production of two KK-gravitons, $\sigma_G = \sigma_{{\rm ve},G} + \sigma_{G G}$, as a function of the DM 
mass $m_{S}$. In all panels, the solid orange (blue dashed) lines represent the total cross-section 
including (not including) mixed KK-graviton production and diagram (c) contribution. The latter case corresponds to
Refs.~\cite{Lee:2013bua} and \cite{Rueter:2017nbk}. In order to appreciate the 
difference, we have included in all panels a zoomed plot in linear scale for the range of 
$m_S$ of interest. Left panel: $\Lambda = 1000$ TeV, $m_{G_1} = 400$ GeV; 
middle panel: $\Lambda = 100$ TeV, $m_{G_1} = 1$ TeV; right panel: $\Lambda = 10$ TeV, $m_{G_1} = 4$ TeV.}
 \label{fig:conjunta}
\end{figure}

In Fig.~\ref{fig:conjunta} we eventually show the total contribution of KK-gravitons to $\left\langle \sigma v \right\rangle$, 
summing virtual KK-graviton exchange and KK-graviton direct production with contributions from the three diagrams in 
Fig.~\ref{fig:diagramas}, $\sigma_G = \sigma_{{\rm ve},G} + \sigma_{G G}$. 
We consider three particular choices of $\Lambda$ and $m_{G_1}$: 
$\Lambda = 1000$ TeV, $m_{G_1} = 400$ GeV (left); $\Lambda = 100$ TeV, $m_{G_1} = 1$ TeV (middle)
and $\Lambda = 10$ TeV, $m_{G_1} = 4$ TeV (right). Our result for $\left\langle \sigma_G v \right\rangle$ is depicted by the solid (orange) line, 
and it is compared with the results shown in the literature (in Refs.~\cite{Lee:2013bua} and \cite{Rueter:2017nbk}), represented by the dashed (blue) line. 
As it can be seen, our results and those in the literature coincide, but for some small differences at large DM masses, $m_{S} \in [1, 6]$ TeV, a range 
shown in the zoomed panel in linear scale.
The net effect of mixed KK-gravitons channels and of diagram (c) in Fig.~\ref{fig:diagramas}
is an increase of the cross-section, that can be as large as a factor two for some specific choices of $\Lambda$ and $m_{G_1}$.
In all panels, the horizontal red dashed line corresponds to the value of the thermally-averaged cross-section for which 
the correct relic abundance is achieved,  $\left\langle \sigma_{\rm FO} v \right\rangle = 2.2 \times 10^{-26}$ cm$^3$/s.
As it was reported in Ref.~\cite{Rueter:2017nbk}, $\left\langle \sigma_{\rm FO} v \right\rangle$
is not achievable through KK-graviton exchange since, even for values of $m_{S}$ such that
$s \sim m_{G_n}^2$, the resonant cross-section is way smaller than
the required one. This result is general and can be found for any value of $\Lambda$ and
$m_{G_n}$, not only for the few examples shown in Fig.~\ref{fig:conjunta}.
On the other hand, as reported in Ref.~\cite{Lee:2013bua}, for larger values of $m_{S}$, 
when the two on-shell KK-graviton production channels take over, a cross-section as 
large as $\left\langle \sigma_{\rm FO} v \right\rangle$ is achievable and the correct relic
abundance can be then reproduced. With respect to Ref.~\cite{Lee:2013bua}, the net effect
of mixed KK-gravitons production and of diagram (c) is to lower slightly the value of 
$m_{S}$ for which $\left\langle \sigma v \right\rangle = \left\langle \sigma_{\rm FO} v \right\rangle$. 
In Fig.~\ref{fig:conjunta}, the red-shaded area represents the theoretical unitarity
bound $\left\langle \sigma v \right\rangle \geq 1/s$, where we can no longer trust the 
theory outlined in Sec.\ref{sec:RS} and higher-order operators should be taken into account.
Notice that, even if in Fig.~\ref{fig:conjunta} the ``untrustable'' region seems to be very near to the value of $m_S$ 
for which the correct relic abundance can be achieved, it is indeed at least
one order of magnitude away, as plots are shown in bi-logarithmic scale.

\begin{figure}[htbp]
\centering
\includegraphics[width=160mm]{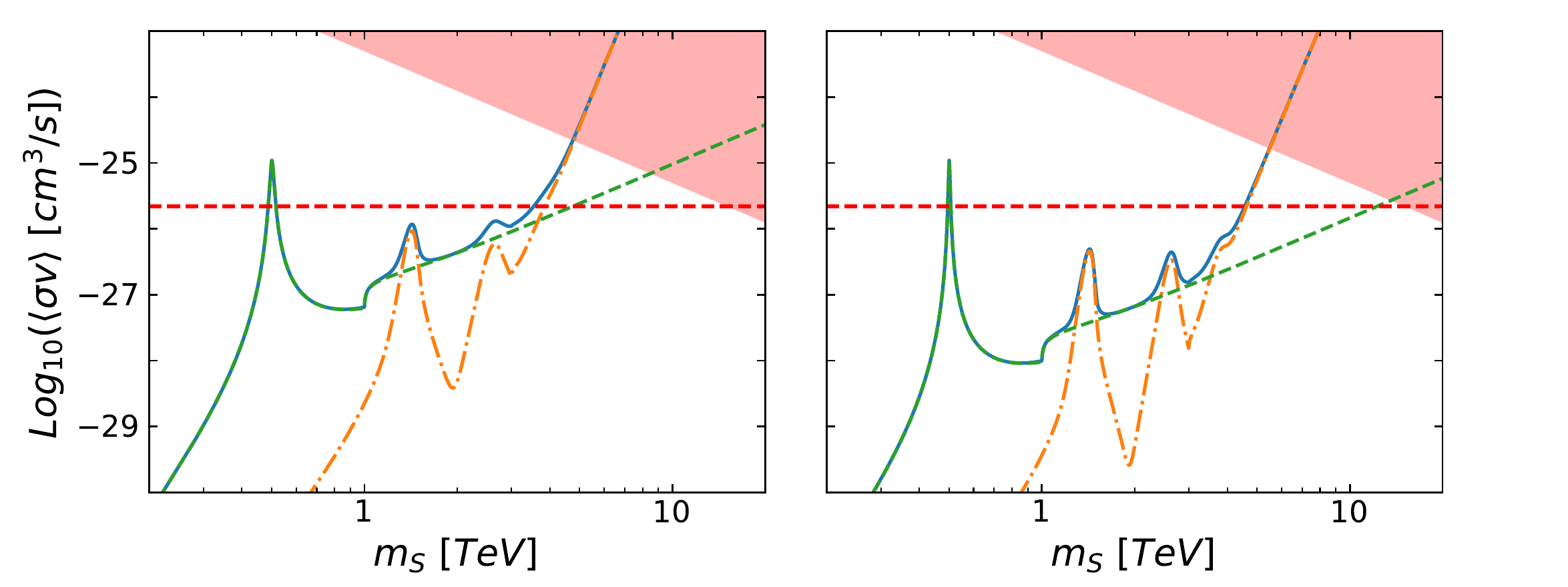}
\caption{\it 
The thermally-averaged scalar DM annihilation cross-section through virtual
radion exchange and direct production of two radions, $\sigma_r = \sigma_{{\rm ve},r} + \sigma_{rr}$ (green, dashed line), as a function of the DM 
mass $m_{S}$, compared with the corresponding cross-section through KK-graviton exchange and production, $\sigma_G$ (orange, dot-dashed line).
The sum of the two cross-sections, $\sigma_r + \sigma_G$, is represented by the (blue) solid line. Left panel:  $\Lambda = 5$ TeV, $m_{G_1} = 3$ TeV and $m_r = 1$ TeV;
Right panel: $\Lambda = 8$ TeV, $m_{G_1} = 3$ TeV and $m_r = 1$ TeV.}
\label{fig:radiones_conjunta}
\end{figure}

\subsection{Virtual radion exchange and on-shell radion production}
\label{sec:radionxsection}

Consider now the case of DM annihilation into SM particles through radion exchange and of direct production of two on-shell radions, 
\be
\sigma_r = \sigma_{{\rm ve},r} (S \, S \rightarrow {\rm SM} \, {\rm SM}) + \sigma_{rr} (S \, S \rightarrow r \, r) \, .
\ee

The analytic expressions for the two relevant radion channels contributing
to $\sigma_r$ can be found in App.~\ref{app:scalarannihilrad}, 
whereas in App.~\ref{app:radion_decay} we give the radion partial decay widths. 
It can be seen that radion decay to fermions is proportional to the  fermion mass  squared, 
$\Gamma (r \to \psi \, \psi) \propto m_r m^2_\psi/\Lambda^2$, whilst radion decay to bosons (either scalar or vector ones) 
is $\Gamma (r \to B \, B) \propto m^3_r/\Lambda^2$. 
Clearly, for radions with ${\cal O}$(TeV) mass bosons decay channels dominate
over fermion ones. However, the decay to massive or massless bosons is rather different: the radion 
decays to photons and gluons at the one-loop level and, therefore, these decay channels are suppressed with 
respect to decays into massive bosons, which proceed at tree level. In summary, the radion decay width is dominated by $r \to WW, r \to ZZ$ and $r \to HH$
(and $r \to SS$ if kinematically possible). 

The two contributions to $\sigma_r$ are shown in Fig.~\ref{fig:radiones_conjunta}, where we plot $\sigma_r$ (green, dashed line) 
as a function of $m_{S}$ and compare it with $\sigma_G$ (orange, dot-dashed line). 
The sum of $\sigma_r$ and $\sigma_G$ is represented by the solid (blue) line. The input parameters for these plots are:
$m_{G_1} = 3$ TeV and $m_r = 1$ TeV; $\Lambda = 5$ TeV (left panel) and $\Lambda = 8$ TeV (right panel).
For these particular choices of $m_{G_1}$, only a couple of KK-graviton resonances appear in $\sigma_G$ before two KK-graviton production takes over.
Again, the red-shaded area represents the theoretical unitarity bound $\left\langle \sigma v \right\rangle \geq 1/s$, where we can no longer trust the 
theory outlined in Sec.\ref{sec:RS}, whilst the red dashed horizontal line is $\left\langle \sigma_{\rm FO} v \right\rangle$.
We can see that, generically and differently from the KK-graviton case, the correct relic abundance can be 
achieved by the resonant virtual radion exchange channel for DM masses around $m_S \sim m_r/2 \left [1 + {\cal O} \left ( m_r^2/\Lambda^2 \right )
\right ]$. 
Since the radion decay width is rather small, for allowed values of $\Lambda$ and radion masses in the TeV range or below,
a significant amount of fine-tuning is needed in order to get the resonant behaviour. In the absence of a theoretical framework 
to explain the specific required relation between $m_{S}$ and $m_r$, we consider difficult to defend this possibility
as an appealing scenario to achieve the observed DM relic abundance. 
On the other hand, as it was the case for the KK-graviton exchange and production shown in Fig.~\ref{fig:conjunta}, the 
target value of  $\left\langle \sigma v \right\rangle$ can be achieved also in the range of DM masses for which 
radion and/or KK-graviton production dominate the cross-section. For the specific values of $m_{G_1}, m_r$ and $\Lambda$ shown 
in Fig.~\ref{fig:radiones_conjunta} this occurs through KK-graviton production. 
We have found that this channel dominates in most of the allowed parameter space, 
while the contribution of radion production is dominant only near the untrustable region $m_{G_1} \sim \Lambda$.

\begin{figure}[htbp]
\centering
\includegraphics[width=160mm]{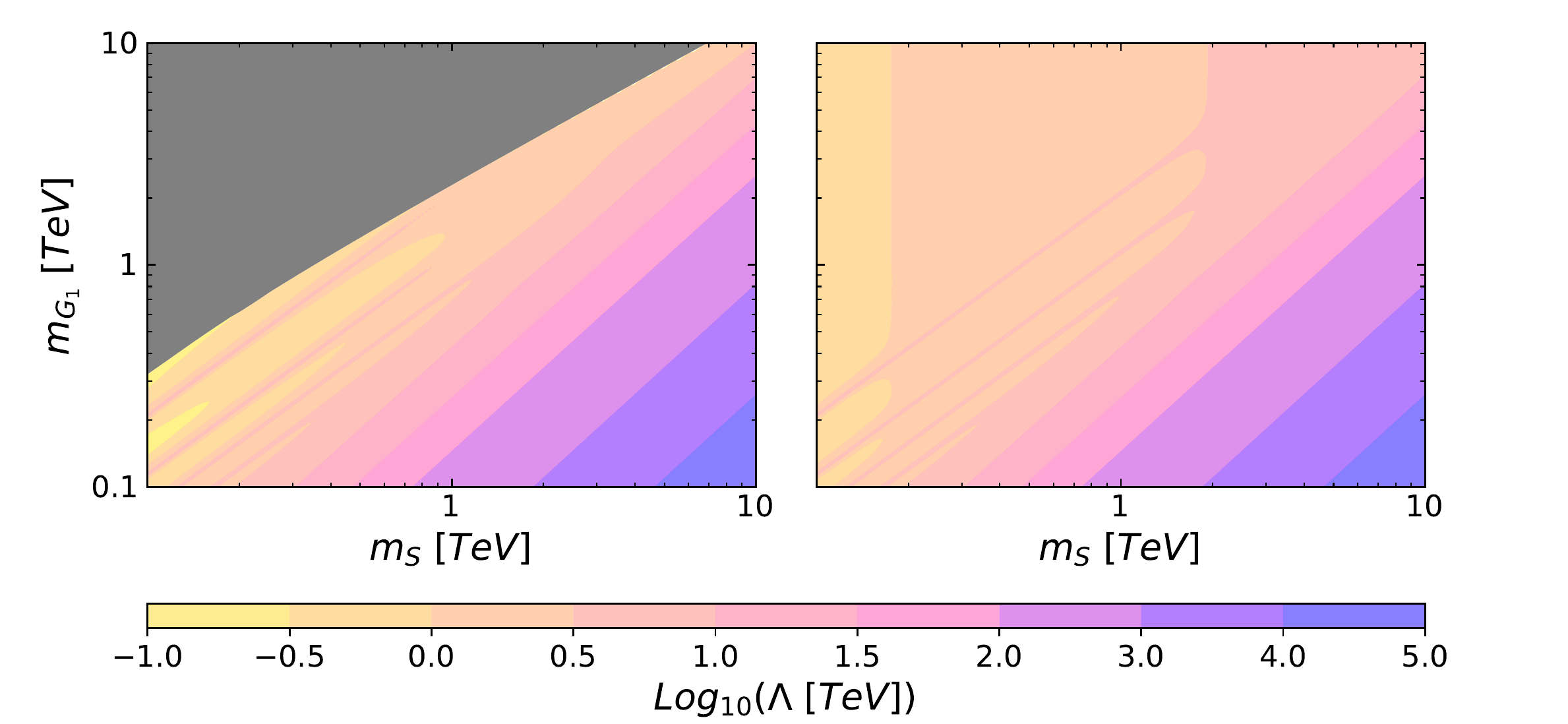}
\caption{\it Values of $\Lambda$ for which the correct DM relic abundance is obtained in the plane ($m_{S}, m_{G_1}$).
Left panel: the extra-dimension length is stabilized without using the radion; Right panel: the extra-dimension length
is stabilized using the Goldberger-Wise mechanism, with a radion mass $m_r = 100$ GeV. The required $\Lambda$ ranges
from 100 GeV to $10^5$ TeV, as shown by the color legend.}
\label{fig:Lambdas}
\end{figure}

In Fig.~\ref{fig:Lambdas} we show the values of $\Lambda$ for which the correct DM relic abundance is obtained in the $(m_{S}, m_{G_1})$ plane. 
In the left panel we assume that the  extra-dimension length is stabilized without introducing  
 the radion field. We can see that $\left\langle \sigma_{\rm FO} v \right\rangle$ can be achieved in a significant
part of the parameter space through KK-graviton production. In order to obtain the target relic abundance
$\left\langle \sigma_{\rm FO} v \right\rangle$ for $m_{S} < m_{G_1}  $, small values of $\Lambda$ are needed, 
usually excluded by LHC data (as we will see in the next section).  Eventually, resonant virtual KK-graviton exchange
is not enough to achieve $\left\langle \sigma_{\rm FO} v \right\rangle$ for $m_{S} \ll m_{G_1}$ for any value of $\Lambda$, 
as it is depicted by the grey region (in agreement with  Ref.~\cite{Rueter:2017nbk}). 

In the right panel we consider, instead, that the extra-dimension length is stabilized using the 
Goldberger-Wise mechanism and we introduce a radion with mass $m_r = 100$ GeV.
In this case, it is always possible to achieve the correct relic abundance: either through resonant radion exchange for $m_S \sim 50$ GeV
(not shown in the plot), through radion production in the region $m_{S} \leq m_{G_1}$ or, for $m_{S} > m_{G_1}$, 
through KK-graviton production.

\section{Experimental bounds and theoretical constraints}
\label{sec:bounds}

As we have seen in Fig.~\ref{fig:Lambdas}, in principle the target relic abundance can be achieved in a 
vast region of the $(m_{S}, m_{G_1})$ parameter space, for $\Lambda$ ranging from $10^{-1}$ TeV to $10^5$ TeV.
However, experimental searches for resonances strongly constrain $m_{G_1}$ and $\Lambda$. 
We will summarize here the relevant experimental bounds and see how only a relatively small region of the parameter space is indeed allowed.

\subsection{LHC bounds}
\label{sec:LHCbounds}

The strongest constraints are given by the resonance searches at LHC. In our model we have considered two types of particles 
that could be resonantly produced at the LHC, the KK-gravitons and the radion.
In order to quantify the impact of LHC data in our parameter space, first of all we need to compute
their production cross-section at the LHC. 


The $n$-th KK-graviton production cross-section at LHC is given by \cite{Giudice:2017fmj}: 
\be
\sigma_{pp \rightarrow G_n}(m_{G_n}) = \frac{\pi}{48 \Lambda^2} 
\left [ 3 {\cal L}_{gg}(m_{G_n}^2) + 4 \sum_q {\cal L}_{q\bar{q}}(m_{G_n}^2) \right ] \, ,
\ee
with 
\be
{\cal L}_{ij}(\hat s) = \frac{\hat s}{s} \int_{\hat s/s}^1 \frac{dx}{x} f_i(x) f_j \left(\frac{\hat s}{x s} \right) \, .
\ee
In our calculations we use the Parton Distribution Functions (PDF's) $f_i(x)$ at $Q^2 = m_{G_n}^2$
obtained from MSTW2008 at leading-order \cite{Martin:2009iq}.


Regarding the radion, since  the $\bar q \, q \, r$ vertex is proportional to the corresponding quark
mass, the production cross-section in $p \, p$ collisions at the LHC is dominated by gluon fusion. 
The gluon-radion interaction is similar to the gluon-Higgs interaction in the SM. We therefore may use the well-known results obtained 
for the SM Higgs production~\cite{Spira:1995rr} rescaling the Lagrangian by a factor $3 v C_3/(2\sqrt{6}\Lambda)$, where $v$ is the standard model VEV. The final expression is given by:
\begin{equation}
\sigma_{pp \rightarrow r}(m_{r}) = \frac{\alpha_s^2 C_3^2}{1536 \pi \Lambda^2} {\cal L}_{gg}(m_{r}^2) \, .
\end{equation}

\begin{figure}[htbp]
\centering
\includegraphics[width=120mm]{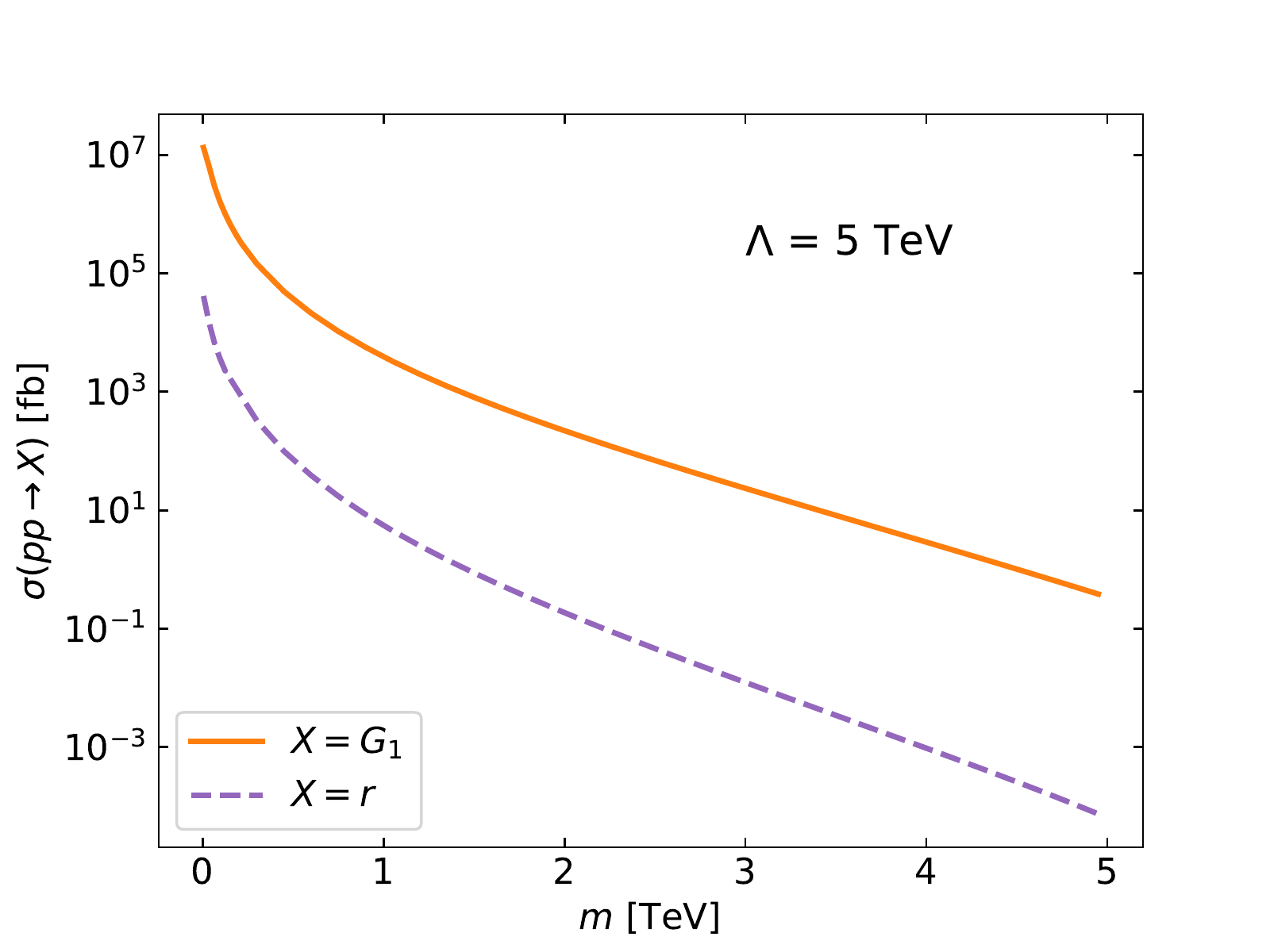}
\caption{\it Theoretical KK-graviton and radion production cross-section at the LHC with $\sqrt{s} = 13$ TeV for $\Lambda = 5$ TeV.}
\label{fig:gravitonproduction}
\end{figure}

In Fig.~\ref{fig:gravitonproduction} we show the production cross-sections for $\Lambda=5$ TeV at $\sqrt{s} = 13$ TeV, where the solid (orange) line stands 
for $p \, p \to G_1$ and the dashed (purple) line for $p \, p \to r$.
It is straightforward to obtain the production cross-sections for a different value of $\Lambda$ by rescaling this plot.
As we can see, the radion production is smaller than graviton production by some orders of magnitude. 
For this reason, the LHC constraints on the Randall-Sundrum model are dominated by (resonant) KK-graviton searches.

\begin{figure}[htbp]
\centering
\includegraphics[width=160mm]{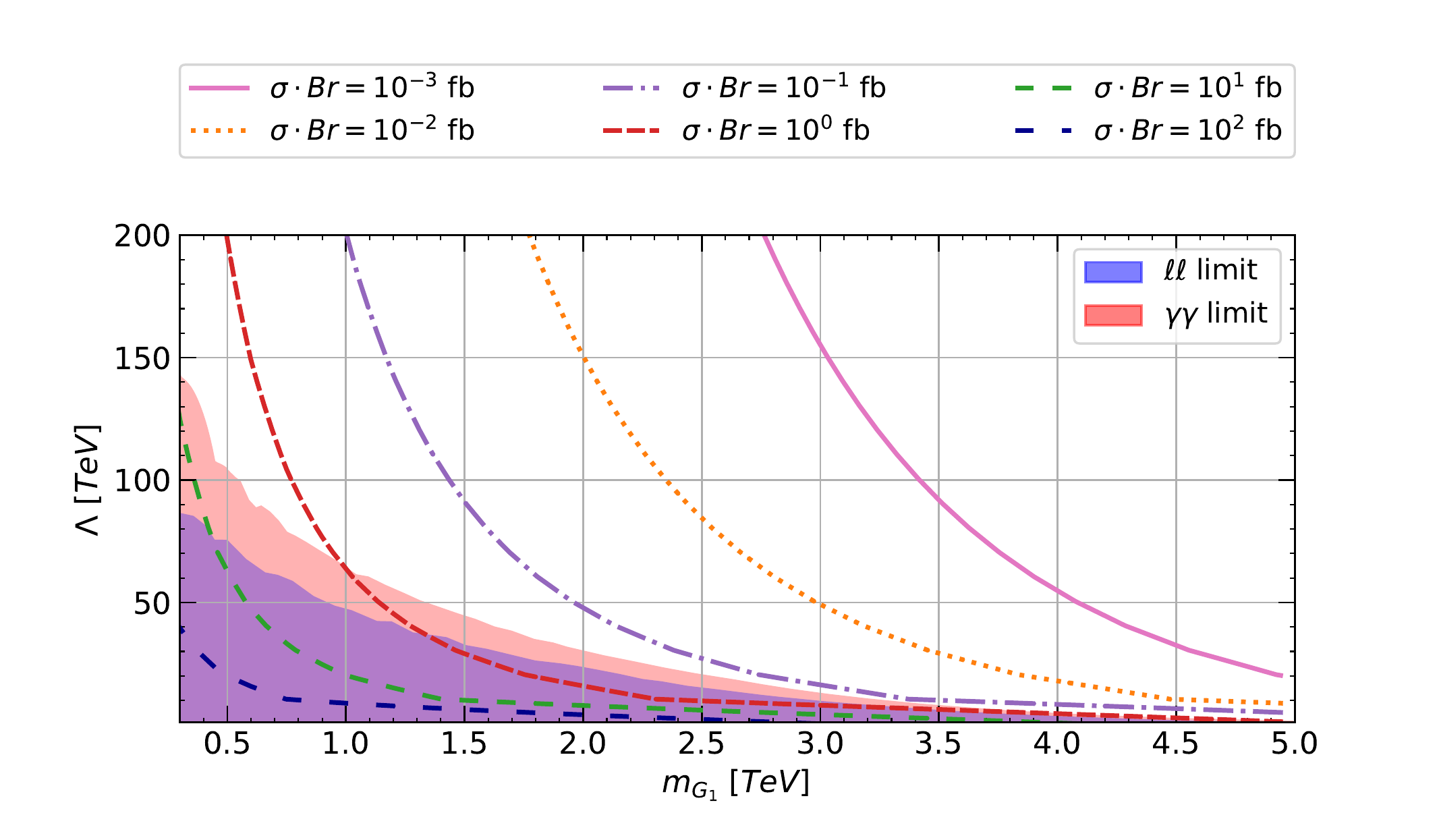}
\caption{\it The exclusion region in the $(m_{G_1}, \Lambda)$ plane 
at the LHC Run II with $\sqrt{s} = 13$ TeV and 36 fb$^{-1}$ through resonant production of KK-graviton
eventually decaying into leptons (light blue) and photons (light red), from Refs.~\cite{Aaboud:2017yyg} and \cite{ATLAS:2017wce}. 
The dashed lines correspond to the functional relation between $\Lambda$ and $m_{G_1}$
for values of $\sigma (p \, p \to G_1) \times {\rm BR}(G_1 \to \ell \, \ell)$ ranging from $10^2$ fb (bottom line) to $10^{-3}$ fb (top line) as in 
the legend.
}
\label{fig:LHCboundsonlambda}
\end{figure}

 The  KK-graviton decay channels that provide the stringest bounds 
on $m_{G_1}$ and $\Lambda$ are $G_1 \rightarrow \gamma\gamma$ \cite{Aaboud:2017yyg} and $G_1 \rightarrow \ell\ell$ \cite{ATLAS:2017wce}. 
In Fig.~\ref{fig:LHCboundsonlambda} we plot the functional dependence over $\Lambda$ and $m_{G_1}$ of
the cross-section $p \, p \to \ell \, \ell$, with $\sigma \times {\rm BR}(G_1 \to \ell \, \ell)$ ranging from $10^2$ fb (bottom line) to $10^{-3}$ fb (top line). 
Comparing the theoretical expectation with the experimental bounds on $\sigma (p \, p \to \ell \, \ell)$ it is possible to draw exclusion regions in 
the $(m_{G_1}, \Lambda)$ plane, given by the darker (blue) shaded area. 
The same can be done using the channel $p \, p \to \gamma \, \gamma$, represented by the lighter (light red) shaded area. 
We can see that the stringest bounds on $\Lambda$ are set by $p \, p \rightarrow G_1 \rightarrow \gamma\gamma$.
Notice that experimental exclusion bounds are given for $m_{G_1} \geq 200$ GeV, approximately. 

\begin{figure}[htbp]
\centering
\includegraphics[width=160mm]{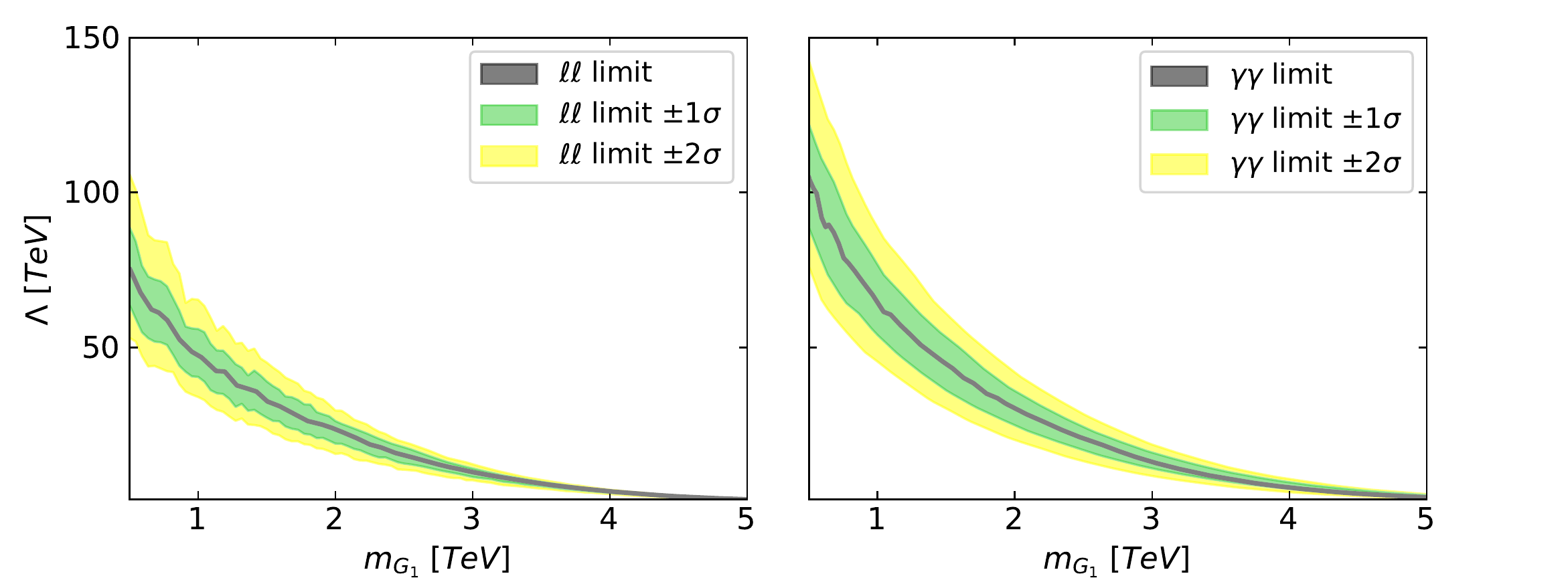}
\caption{\it 
Bounds over $\Lambda$ as a function of $m_{G_1}$ from the LHC with $\sqrt{s} = 13$ TeV and 36 fm$^{-1}$, 
from Refs.~\cite{Aaboud:2017yyg} and \cite{ATLAS:2017wce}. Red and blue lines
represent the 1$\sigma$ and 2$\sigma$ error on the constraint, respectively. The resonance (to be understood as the first KK-graviton mode) 
 eventually decays into leptons (left panel) or  into photons (right panel). 
}
 \label{fig:LHC}
\end{figure}

In Fig.~\ref{fig:LHC} we show the statistical uncertainties on the experimental bound on $\sigma (p \, p \to \ell \, \ell)$ (left panel) and 
$\sigma (p \, p \to \gamma \, \gamma)$ (right panel), where the yellow and green bands are the bounds at 1$\sigma$ and 2$\sigma$
in the $(m_{G_1},\Lambda)$ plane, respectively.
It can be seen that for low KK-graviton mass the bounds on $\Lambda$ suffer from a large indetermination:
in this range we can only say that $\Lambda$ should be larger than some value ranging from 50 to 100 TeV, approximately. 

\subsection{Direct and Indirect Dark Matter Detection}
\label{sec:direct}

\begin{figure}[htbp]
\centering
\begin{tabular}{cc}
\includegraphics[width=160mm]{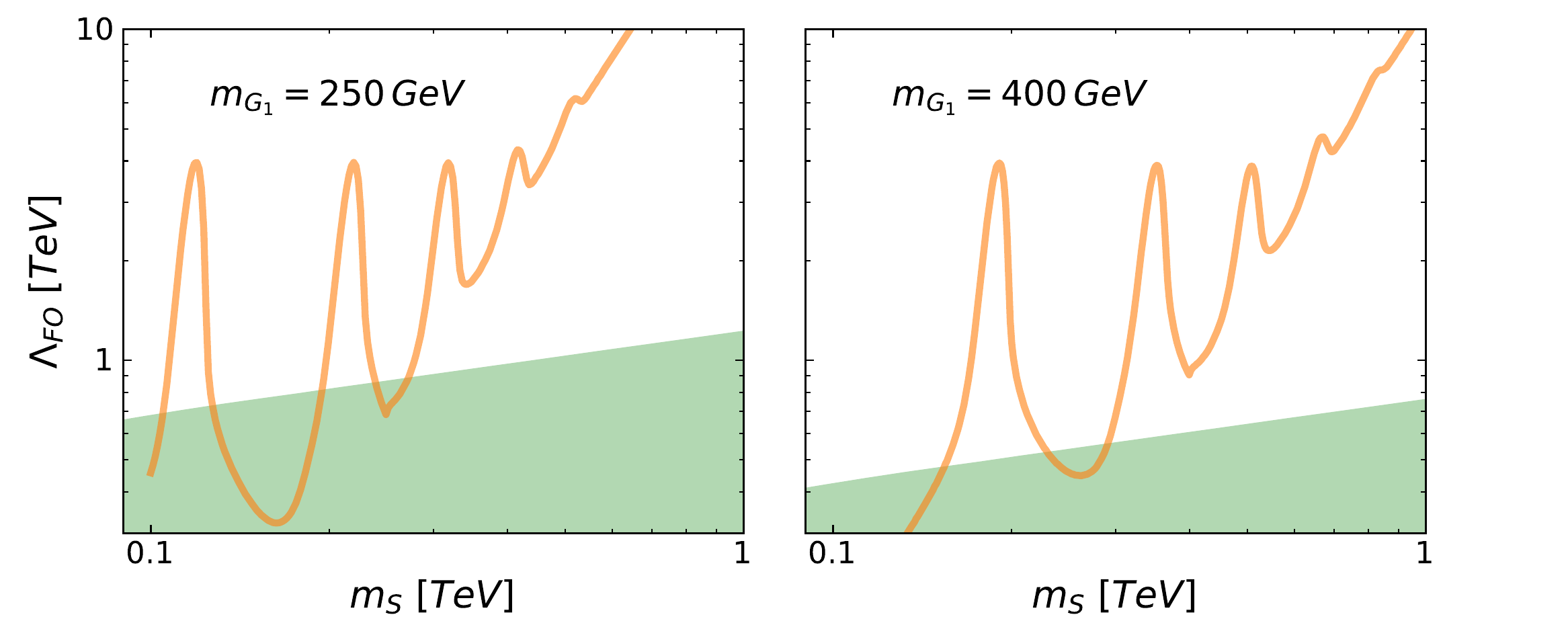}
\end{tabular}
\caption{\it The DD bounds in the ($m_S, \Lambda$) plane for two values of $m_{G_1}$, represented by the green-shaded area. 
Also shown is the dependence of $\Lambda_{\rm FO}$ on the scalar DM mass $m_S$ for fixed $m_{G_1}$,
being $\Lambda_{\rm FO}$ the value of $\Lambda$ for which the freeze-out thermally-averaged cross-section $\langle \sigma_{\rm FO} v \rangle$ 
is achieved for the chosen values of $m_S$ and $m_{G_1}$. 
Left panel: $m_{G_1} = 250$ GeV; Right panel: $m_{G_1} = 400$ GeV.
}
\label{fig:LambdadepatFO}
\end{figure}

The total cross-section for spin-independent elastic scattering between dark matter and nuclei reads \cite{Carrillo-Monteverde:2018phy}:
\be
\sigma_{{\rm DM}-p}^{\rm SI} = \left [ \frac{m_p \, m_{S} }{A \pi (m_{S} + m_p)} \right ]^2\left [ Af_p^{S} + (A-Z)f_n^{S}  \right ]^2 \ , 
\label{eq:DD}
\ee
where $m_p$ is the proton mass, while  Z and A are the number of protons and the atomic number.
The nucleon form factors are given by 
\be
\left \{
\begin{array}{lll}
f_p^{\rm DM} &=& \frac{m_{S} \, m_p}{4 m_{G_1}^2\, \Lambda^2} 
\left \{ \sum_{q=u,c,d,b,s} 3 \left [ q(2) + \bar{q}(2) \right ] + \sum_{q=u,d,s} \frac{1}{3}  f_{Tq}^p \right \} \, ,\\
&& \\
f_n^{\rm DM} &=& \frac{m_{S} \, m_p}{4 m_{G_1}^2 \, \Lambda^2} 
\left \{ \sum_{q=u,c,d,b,s} 3 \left [ q(2) + \bar{q}(2) \right ] + \sum_{q=u,d,s} \frac{1}{3} f_{Tq}^n \right \} \, ,
\end{array}
\right .
\ee
with $q(2)$  the second moment of the quark distribution function
\be
q(2) = \int_0^1 dx \ x \ f_q(x)
\ee
and $f_{Tq}^{N = p, n}$  the mass fraction of light quarks in a nucleon: $f^p_{Tu} = 0.023$, $f^p_{Td} = 0.032$ and $f^p_{Ts} = 0.020$ for a proton and $f^n_{Tu} = 0.017$, $f^n_{Td} = 0.041$ and $f^n_{Ts} = 0.020$ 
for a neutron \cite{Hisano:2010yh}. The strongest bounds from Direct Detection (DD) Dark Matter searches are found at
 XENON1T, which uses as  target mass $^{129}$Xe, ($Z=54$ and $A-Z=75$). 
In order to compute the second moment of the PDF's we have used Ref.~\cite{Martin:2009iq}  and the exclusion curve of 
XENON1T~\cite{Aprile:2017iyp} to set constraints on the $(m_S, m_{G_1}, \Lambda)$ parameter space. 
Our results are shown in Fig.~\ref{fig:LambdadepatFO}, where we depict the DD bounds in the ($m_S, \Lambda$) plane for two values of $m_{G_1}$, 
$m_{G_1} = 250$ GeV (left panel) and $m_{G_1} = 400$ GeV (right panel). 
Also shown is the dependence of the value of $\Lambda$ required to achieve the observed relic abundance, $\Lambda_{\rm FO}$, 
as a function of the scalar DM mass $m_S$.  
The resonant behaviour of $\Lambda_{\rm FO}$ for different values of $m_S$ shows that, for low values of $m_S$ and $m_{G_1}$, 
the cross-section is dominated by virtual KK-graviton exchange. For larger values of $m_S$ at fixed $m_{G_1}$ production of KK-gravitons 
takes over and $\Lambda_{\rm FO}$ grows smoothly with $m_S$. 
The region of the parameter space excluded by DD experiments is represented by the green-shaded area at the bottom of the two plots. 
Due to the fact that in the excluded region the dominant channel to achieve $\langle \sigma_{\rm FO} v \rangle$ is KK-graviton virtual exchange, the 
exclusion bounds will show a characteristic striped pattern (as it will be shown in Fig.~\ref{fig:Completa}). 
We have found, however, that constraints from DD experiments are always much weaker than those obtained at the LHC.

Regarding  DM indirect searches, there are several experiments looking for astrophysical signals: 
for instance, the Fermi-LAT collaboration has analyzed the gamma ray flux 
arriving at the Earth from Dwarf spheroidal galaxies \cite{Fermi-LAT:2016uux} and  the galactic center 
\cite{TheFermi-LAT:2015kwa,TheFermi-LAT:2017vmf}, while AMS-02 has reported data about the positrons 
\cite{PhysRevLett.113.221102} and antiprotons \cite{PhysRevLett.117.091103} coming from the center of the galaxy.
These results are relevant for DM models  that generate a continuum spectra of different SM particles, such as the RS scenario we are considering. 
Recall that DM annihilation into a pair of SM particles via KK-graviton exchange is $d$-wave--suppressed and, therefore, only the annihilation channels into either KK-gravitons or radions lead to observable signals.
 Both of them will then decay into SM particles leading to a continuum spectrum
 \footnote {We disregard the fine-tuned possibility of achieving the target DM relic density via resonant radion exchange, as discussed in SSec.ec. \ref{sec:Results}.}. 
However, current data from indirect detection experiments allows to constrain DM masses below $\sim$ 100 GeV (provided the annihilation cross-section is not velocity suppressed), while for our case of heavy DM 
($\sim$ 1 TeV) the limits on the cross-section are well above the required value $\langle \sigma_{\rm FO} v \rangle$. 
Thus, indirect searches have no impact on the viable parameter space (see however Ref. \cite{Lee:2014caa} for other DM scenarios based on RS).

\subsection{Theoretical constraints}
\label{sec:unitarityandOPE}

Besides the experimental limits, there are mainly two theoretical concerns about the validity of our calculations which affect part of the 
$(m_S, m_{G_1},\Lambda)$ parameter space. The first one is related to the fact that we are performing just a tree-level computation of the relevant DM annihilation cross-sections, and we should worry about unitarity issues. In particular, the t-channel annihilation cross-section into a pair of 
KK-gravitons, $\sigma(S S \to G_n G_m)$, diverges as $m_S^8/(m_{G_n}^4 m_{G_m}^4)$ 
in the non-relativistic limit $s \simeq m_S^2$, so it is important to check that the effective theory is still unitary. We estimate the unitarity bound as 
$ \sigma < 1/s \simeq 1/m_S^2$,  showing as a green-meshed area in Fig.~\ref{fig:Completa} the region in which such bound is not satisfied and 
therefore our calculation is not fully reliable.

The second theoretical issue refers to the consistency of the effective theory framework: in the Randall-Sundrum scenario, at energies somewhat larger than $\Lambda$
the KK-gravitons are strongly coupled and the five-dimensional field theory from which we start is no longer valid. We therefore impose that at least $m_{G_1} < \Lambda$ to trust  our results\footnote{We will see that, in the allowed region, also the relation $m_S < \Lambda$ is fulfilled.}. 
 Notice that this constraint is general for any effective field theory: since we are including the first KK-gravitons in the low energy spectra, for the effective theory to make sense the cut-off scale $\Lambda$  should be larger  than the masses of such states.

\section{Achieving the DM relic abundance in RS}
\label{sec:Results}

We show in this section the allowed parameter space for which the target
value of $\left\langle \sigma v \right\rangle$ needed to achieve the correct DM relic abundance in the freeze-out scenario, 
($\left\langle \sigma_{\rm FO} v \right\rangle = 2.2 \times 10^{-26}$ cm$^3$/s) can be obtained, 
taking into account both the experimental bounds and the theoretical constraints outlined in Sec.~\ref{sec:bounds}.

\begin{figure}[htbp]
\centering
\includegraphics[width=160mm]{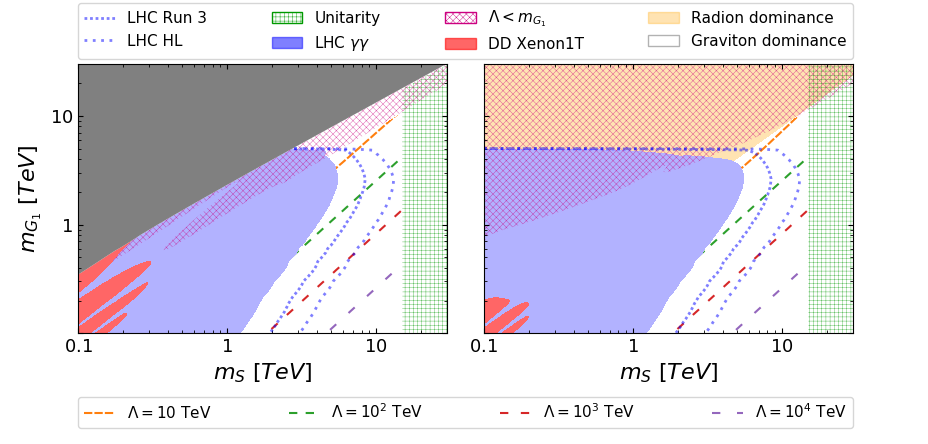}
\caption{\it Region of the $(m_{S}, m_{G_1})$ plane for which 
$\left\langle \sigma v \right\rangle = \left\langle \sigma_{\rm FO} v \right\rangle$. 
Left panel:  the radion and the extra-dimension stabilization mechanism play no role 
in DM phenomenology.
Right panel:  the extra-dimension length is stabilized with  the Goldberger-Wise mechanism, 
with radion mass $m_r = 100$ GeV.
In both panels, the grey area represents the part of the parameter
space where it is impossible to achieve the correct relic abundance; 
the red-meshed area is the region for 
which the low-energy RS effective theory is untrustable, as $\Lambda < m_{G_1}$; 
the wiggled red area in the lower left corner is the region excluded by DD experiments; 
the blue area is excluded by resonant KK-graviton searches at the LHC with 36 fb$^{-1}$ at $\sqrt{s} = 13$ TeV; 
the dotted blue lines represent the expected LHC exclusion bounds at the end of the Run III (with $\sim 300$ fb$^{-1}$)
and at  the HL-LHC (with $\sim 3000$ fb$^{-1}$); eventually, the green-meshed area on the right is
the region where the theoretical unitarity constraints are not fulfilled.
In the left panel, the allowed region is represented by the white area, for which $\left\langle \sigma_{\rm FO} v \right\rangle$
is obtained through on-shell KK-graviton production. In the right panel, in addition to the white area, 
within the tiny orange region  $\left\langle \sigma_{\rm FO} v \right\rangle$
is obtained through on-shell radion production and virtual radion exchange.
The dashed lines depicted in the white region represent the values of $\Lambda$ needed to obtain the correct relic abundance
(as in Fig.~\ref{fig:Lambdas} of Sec.~\ref{sec:annihilres}).
}
\label{fig:Completa}
\end{figure}

Our final results are shown in Fig.~\ref{fig:Completa}, where we draw the allowed regions 
of the $(m_{S}, m_{G_1})$ plane for which $\left\langle \sigma v \right\rangle = \left\langle \sigma_{\rm FO} v \right\rangle$. 
In the left panel, we are agnostic about the extra-dimension length stabilization mechanism, 
and assume that neither the unspecified  mechanism nor the radion have an impact on  the DM phenomenology, as would be the 
case for instance if all the new particles in this sector are heavier than the TeV scale; 
in the right panel, we take into account the radion and consider  the Goldberger-Wise mechanism to  stabilize the extra-dimension length. 
The radion mass in this case can be somewhat smaller than the TeV scale (see Sec.~\ref{sec:rad}), 
and therefore it can be relevant for DM annihilation, as we will discuss below.
We show our findings for $m_r = 100$ GeV,  but other values of $m_r$ lead to similar results. 
As a guidance, the dashed lines taken from Fig.~\ref{fig:Lambdas} represent the values of $\Lambda$ needed to achieve the relic abundance in a particular point of the $(m_S, m_{G_1})$ plane. 
The color legend for the two plots is given in the Figure caption.

\subsection{KK-graviton contributions}
\label{sec:graviton}

Let's consider first the case in which the relic abundance is obtained through virtual KK-graviton exchange
and/or on-shell KK-graviton production (left panel). We can distinguish two regions of the parameter space: 
\begin{enumerate}
\item $m_{G_1} > m_{S}$ \\ 
In this regime the DM annihilates via KK-graviton exchange to SM particles, only. 
As we have seen in Fig.~\ref{fig:contribuciones}, the annihilation cross-section is rather small. The grey shaded area in the plot
represents the region of the ($m_S, m_{G_1}$) plane for which it is not possible to get $\left\langle \sigma_{\rm FO} v \right\rangle$.
Below this region, in principle we could find a value of $\Lambda$ low enough to reach the target relic abundance
via resonant KK-graviton exchange.  This is, however, in conflict with exclusion bounds in the 
($m_{G_1},\Lambda$) plane from LHC (see Fig.~\ref{fig:LHCboundsonlambda}), represented by the darkest (blue) shaded area 
In addition to the stringent LHC Run II bounds, 
if the $\Lambda$ needed to achieve $\left\langle \sigma_{\rm FO} v \right\rangle$ 
for a given $m_S$ is smaller than $m_{G_1}$, we can no longer trust the RS model as a viable effective low-energy formulation of gravity
(diagonal red-meshed area).
Therefore, due to the combination of experimental bounds and theoretical constraints, 
for $m_{G_1} > m_{S}$ is not possible to obtain $\left\langle \sigma_{\rm FO} v \right\rangle$, 
as it was indeed found in Ref.~\cite{Rueter:2017nbk}.

\item $m_{G_1} < m_{S}$ \\
In this case, although the $S \, S \to {\rm SM} \, {\rm SM}$ channel is still open,  the target cross-section is achievable
through production of on-shell KK-gravitons, $S \, S \rightarrow G_n \, G_m$. Due to the LHC Run II bounds, the region 
of the ($m_{G_1},\Lambda$) plane for which we can obtain $\left\langle \sigma_{\rm FO} v \right\rangle$ corresponds mainly
to the region for which $(m_{G_1}/m_{S})^2 \ll 1$. In this region, the value of $\Lambda$ needed to reach the freeze-out
relic abundance is in the range $\Lambda \in [10, 10^4]$ GeV, in agreement with the stringent LHC Run II bounds on $\Lambda$
for relatively low $m_{G_1}$. At large values of $m_S$ the theoretical unitarity bound discussed in Sec.~\ref{sec:unitarityandOPE} is relevant
and, therefore, $m_S$ cannot be much larger than 10 TeV (vertical green-meshed area). Eventually, the white area represents
the region of the parameter space for which the freeze-out scenario can produce the correct DM relic abundance. 
Notice that most of this region could be tested either by the LHC Run III\footnote{This region could be already partially tested using the 
complete LHC Run II analysis, with 100 fb$^{-1}$, not included in this paper.} (with expected 300 fb$^{-1}$)
or by the High-Luminosity LHC (with nominal 3000 fb$^{-1}$), as shown by the dotted lines depicted in the Figure.
Typical values for $m_S, m_{G_1}$ and $\Lambda$ in the region that would still be allowed after HL-LHC are
$m_S \in [3,15]$ TeV, $m_{G_1} < 1$ TeV and $\Lambda > 10^3$ TeV (although a tiny region around $m_S \sim 10$ TeV with 
$m_{G_1}$ as large as few TeV with $\Lambda \in [10,100]$ TeV could also be viable).
\end{enumerate} 


The wiggled dark shaded (red) region in the lower left corner is the bound imposed by XENON1T. The peculiar shape of the bound is a consequence of the resonances in the DM annihilation channels via virtual graviton exchange (see Fig.~\ref{fig:LambdadepatFO}). We can see that the DD bounds are much weaker than those from the LHC.

\subsection{Radion contribution}
\label{sec:radion}

Let's consider now the case in which, in addition to virtual KK-graviton exchange
and/or on-shell KK-gravitons production, DM could also produce virtual or real radions (right panel). 
To make easy the comparison with the previous situation, we again consider two regimes:
\begin{enumerate}
\item $m_{G_1} > m_{S}$ \\ 
It is always possible to achieve the correct relic abundance through resonant virtual radion exchange and on-shell radion production (see Fig.~\ref{fig:radiones_conjunta}). 
In the right plot of Fig.~\ref{fig:Completa} the former would occur for $m_S = 50$ GeV, outside the range depicted in the Figure.
Being the radion width extremely narrow, this is possible only in presence of 
a significant fine-tuning of the DM mass $m_S$ and of the radion mass, $2 m_S \sim m_r$. In the absence of a theoretical motivation 
for such a relation between two, in principle, uncorrelated parameters, we consider this mechanism to achieve the target relic abundance
not {\em natural}. In the region considered in the plot, the relic abundance can be also achieved through production of on-shell radions 
for very low values of $\Lambda$. 
This region is represented by the orange (lightest) shaded area. Most of this region, however, is excluded when asking 
 $\Lambda$ to be larger than $m_{G_1}$, as one can see by  the diagonal red-meshed area in the plot, $\Lambda < m_{G_1}$.
After taking into account the LHC Run II bounds and the limit of validity of the RS model as an effective low-energy theory, a tiny orange-shaded 
region at $m_S \sim 4$ TeV, $m_{G_1} \sim 5$ TeV and $\Lambda \in [5, 10]$ TeV is still not excluded. Most of it will 
be tested with the LHC Run III. 
\item $m_{G_1} < m_{S}$ \\
Since the real KK-graviton production channel, once kinematically open grows very fast as $(m_{S}/m_{G_1})^8 $ (see Fig.~\ref{fig:radiones_conjunta}), 
it easily dominates the cross-section. Therefore, in  this region of the parameter space
there are no significant differences with respect to the case in which the radion is absent, discussed 
in Sec.~\ref{sec:graviton}.

\end{enumerate}

\subsection{Remarks about other setups}

In this paper we have focused on the original RS model, in which all the SM particles (and also the DM in our case) are localized on the IR-brane.
In the absence  of graviton brane localized kinetic terms (BLKT's), within this setup all the SM and DM fields couple to the full tower of KK-graviton 
excitations with universal strength, $\Lambda^{-1}$.
As we have seen, the strong bounds  from LHC Run II lead to quite large allowed values of $\Lambda$ ($\gtrsim 10$ TeV), which 
somehow reintroduce a little hierarchy problem.
However many other different configurations have been studied, allowing for some of (or all)   the SM fields to propagate in the bulk;
 for instance, placing gauge bosons and fermions in the bulk has the potential to also explain the hierarchy of fermion masses.
Moreover, these extra-dimensional scenarios can be  interpreted as strongly-coupled models in four dimensions (see Ref.~\cite{Lee:2013bua}
 for details of this duality). 
 
 Several of the above possibilities have been already analyzed in the context of gravity-mediated DM that we are addressing, 
 including DM candidates of various spins (0,1/2 and 1).
 The idea is that  the propagation of SM fields in the bulk and the introduction of BLKT's  can reduce suitably the coupling of the SM 
 particles to the KK-gravitons, relaxing the LHC bounds and allowing for lower values of $\Lambda$ which would then satisfy the original motivation of RS 
 models for solving  the hierarchy problem. Although to study in detail these alternative RS scenarios is beyond the scope of this paper, 
 we want to comment in this section about the impact of our results on such other models.
 
 In Ref.~\cite{Rueter:2017nbk}, besides the scenario considered here with all SM and DM fields 
 localized in the IR-brane, two additional benchmark models were studied: 1) SM gauge bosons in the bulk with third generation quarks confined in the IR brane, 
 and all other SM fermions localized close to the UV-brane, so that their couplings to the KK-graviton modes are negligible, and 
 2) SM fermions localized at various places in the bulk to explain the observed fermion masses and SM gauge bosons propagating also in the bulk.
 In all scenarios, the Higgs field should remain close to the IR-brane to solve the hierarchy problem, and the DM is also assumed to be 
 localized on the IR-brane. While in none of these setups it was possible to obtain the correct relic density for scalar DM through virtual KK-graviton exchange, the authors did not consider the annihilation channel $ S S \to G_n G_m$ nor $ S S \to r r $.
Since these channels will occur with the same cross-section as in the IR-brane model we analyzed in this paper, 
it is clear that also in the cases considered in Ref.~\cite{Rueter:2017nbk} it would be possible to get the target value 
$\left\langle \sigma_{\rm FO} v \right\rangle $ when  $m_S > m_{G_1}$. Actually, it would be easier than in the case considered here, 
as the LHC bounds on $\Lambda$ are weaker.

In Ref.~\cite{Lee:2014caa} two additional setups where analyzed and also confronted with indirect bounds from astrophysical data: 
model A, which addresses the hierarchy problem  with the Higgs and DM localized on the IR-brane and the SM matter on the UV-brane, 
and model B (that gives up the hierarchy problem) where only DM is localized on the IR-brane while the SM matter and Higgs fields are 
confined to the UV-brane. In both cases, SM gauge bosons propagate in the bulk, so that there is a hierarchy of couplings of the KK-graviton modes,  
being of order $\Lambda^{-1}$ for DM (and the Higgs field in model A) but conveniently suppressed for gauge bosons and 
negligible for SM matter  fields (and the Higgs in model B).
As a consequence, the standard radion and KK-graviton searches at  LHC do not apply to these models and other searches 
should be re-interpreted to obtain bounds. Therefore, much lower values of $\Lambda$ and $m_{G_1}$ would still be allowed and
it should be possible to achieve the correct relic abundance for DM masses  in a wider range, from few GeV to TeV, 
in agreement with our results in Fig.~\ref{fig:Lambdas}. 
 
In the dual picture of the RS model, the radion is dual to the dilaton, the Goldstone boson from dilatation symmetry in 4D.
The dilaton couplings  
are fixed by scale invariance, and turn out to have the same structure as the radion couplings at linear order. 
In Refs.~\cite{Bai:2009ms,Blum:2014jca}, the case in which DM couples to the SM only through a dilaton was studied
The authors found that the correct relic abundance can be achieved for light dilaton and DM, since collider bounds from dilaton searches 
are weaker than for the KK-graviton modes (the dilaton production cross-section is about two - three orders of magnitude smaller 
than the KK-graviton one, as we can see in Fig.~\ref{fig:gravitonproduction}). However, as we are studying a consistent gravitational theory
and not only the SM plus a dilaton field, the much stringent bounds from KK-gravitons searches do apply.

\section{Conclusions}
\label{sec:concl}

In this paper we have explored the possibility that the observed Dark Matter component in the Universe is represented by some new scalar particle with a mass in the TeV range. This particle interacts with the SM particles only gravitationally (in agreement with non-observation of DM signals at both direct and indirect detection DM 
experiments). Although this hypothesis would, in principle, mean that the interaction with SM particles is too feeble to reproduce the observed DM relic abundance, we show that this is not the case once this setup is embedded in a warped
extra-dimensional space-time, along the ideas of the Randall-Sundrum proposal of Ref.~\cite{Randall:1999ee}. 
We consider, therefore, two 4-dimensional branes in a 5-dimensional AdS$_5$ space-time at a separation $r_c$, very small compared 
with present bounds on deviations from Newton's law. On one of the branes, the so-called ``IR-brane", both the SM particles and a
scalar DM particle are confined, with no particle allowed to escape from the branes to explore the bulk.
In this particular extra-dimensional setup, gravitational interaction between particles on the IR-brane, in our case between a scalar DM particle
and any of the SM particles, occurs with an amplitude proportional to $1/M^2_P$ when the two particles exchange a graviton zero-mode, 
but with a suppression factor $1/\Lambda^2$ when they do interact exchanging higher KK-graviton modes. Since $\Lambda$ can 
be as low as a few TeV (due to the warping effect induced by the curvature of the space-time along the brane separation), clearly a huge enhancement of the cross-section is possible with respect to standard linearized General Relativity.

Using this mechanism, it was studied in the literature if the observed relic abundance in the Universe can be obtained
through resonant KK-graviton exchange via $\sigma ({\rm DM} \, {\rm DM} \to G_n \to {\rm SM} \, {\rm SM})$
(for any spin of the DM particle),  showing that taking into account the LHC bounds on $\Lambda$ as a function of the mass of the first KK-graviton, $m_{G_1}$, it is impossible to achieve the target value of the thermally-averaged cross-section $\langle \sigma_{\rm FO} \, v \rangle $ 
for any value of $m_{\rm DM}$ if the DM particle has spin 0 or 1/2 \cite{Rueter:2017nbk}. 
In Refs.~\cite{Lee:2013bua,Lee:2014caa,Han:2015cty,Carrillo-Monteverde:2018phy} it was however shown that, 
for DM masses larger then the KK-graviton mass, another annihilation channel opens, namely
DM annihilation into two (identical) KK-gravitons, $\sigma ({\rm DM} \, {\rm DM} \to G_n \, G_n)$. 
In this paper, we have studied the possibility that this channel may give a cross-section large enough to attain the
observed relic abundance, for the particular case of a scalar DM particle with mass $m_S$.
We have indeed found that this is the case and that the region of the parameter space
for which $\langle \sigma \, v \rangle \sim \langle \sigma_{\rm FO} \, v \rangle $ is typically at $m_S$ of the order of a few TeV, 
compatible with present direct production searches at the LHC.
In the references above some effects were overlooked, though. In particular, a quadratic interaction of the DM particles with KK-gravitons
({\em i.e.} the existence of a $S \, S \, G_n \, G_m$ vertex when expanding the metric up to second order about the Minkowski metric)
was not considered. This amplitude is of the same order in $1/\Lambda$ as the t- and u-channel contributions to 
$\sigma ({\rm DM} \, {\rm DM} \to G_n \, G_n)$ considered in the literature and, 
by increasing the cross-section at large value of the DM mass, lowers the value of $m_S$ needed to achieve
the relic abundance at fixed value of $m_{G_1}$. The same effect is also induced by the possibility of the DM particles
annihilating into different KK-gravitons, $\sigma ({\rm DM} \, {\rm DM} \to G_m \, G_n)$, something allowed since translational
invariance along the 5-th dimension is explicitly broken by the presence of the branes. This was also overlooked in the
existing literature. These effects and their impact have been discussed extensively in Sec.~\ref{sec:annihilres} and App.~\ref{app:annihil}.

After having computed the relevant contributions to the cross-section, we have scanned the parameter space of the model 
(represented by $m_S$, $m_{G_1}$ and $\Lambda$), looking for regions in which the observed relic abundance can be achieved. 
This region has been eventually compared with experimental bounds from resonant searches at the LHC Run II and from 
direct and indirect DM detection searches, finding which portion of the allowed parameter space is excluded by data. 
Eventually, we have studied the theoretical unitarity bounds on the mass of the DM particle and on the validity of the RS model
as a consistent low-energy effective theory. Our main result is that a significant portion of the $(m_S,m_{G_1})$ plane where $m_S > m_{G_1}$ can reproduce the observed relic abundance, for values of $\Lambda$ ranging from a few to thousands of TeV and $m_S \in [1,10]$ TeV. Unitarity bounds put a (theoretical)
upper limit on the mass of the DM particle and, interestingly enough, most part of the allowed parameter space could therefore 
be tested by the LHC Run III and by the proposed High-Luminosity LHC. 

In the presence of a Goldberger-Wise mechanism to stabilize the separation between the two branes,
the radion $r$ is expected to be light, $m_r \lesssim {\cal O}$(TeV), and DM can also
annihilate into SM particles via the exchange of a virtual radion and, for $m_S > m_r$, two DM particles can also produce
directly two on-shell radions. This has been studied in detail in Sec.~\ref{sec:radionxsection} and App.~\ref{app:scalarannihilrad}.
Since, contrary to the KK-graviton mass (strongly related to $\Lambda$ in the RS setup), the radion 
mass is in practice a free parameter of the model (depending on the unknown details of the scalar potential in the bulk and of
some brane-localized terms), it is possible to achieve $\langle \sigma_{\rm FO} \, v \rangle $ for any value of $m_S$ and 
$m_{G_1}$, even in the case $m_{G_1} > m_S$, through the resonant radion exchange channel (at the price
of introducing a significant, theoretically unappealing, fine-tuning of the DM mass with respect to the radion mass, $2 m_S \sim m_r$)
or through on-shell radion production. The region for $m_{G_1} > m_S$, however, is mostly
excluded due to the fact that the value of $\Lambda$ needed to reach the target relic abundance is $\Lambda < m_{G_1}$, a condition
that makes untrustable the RS model as a valid effective low-energy theory. Apart from a tiny region for which the two radion on-shell production 
channel dominates in the cross-section, the rest of the allowed parameter space is similar to that found in the absence of a radion.

 \section*{Acknowledgements} 

We thank, Hyun Min Lee, Myeonghun Park and Ver\'onica Sanz for correspondence about the DM annihilation cross-sections into KK-gravitons and radion interactio. This work has been partially supported by the European Union projects H2020-MSCA-RISE-2015 and H2020-MSCA- ITN-2015//674896-ELUSIVES and 
by the Spanish MINECO under grants FPA2014-57816-P, FPA2017-85985-P and  SEV-2014-0398.

\appendix

\section{Spin 2 massive graviton}
\label{app:spin2}

The propagator of the $n$-th KK-graviton mode, with mass $m_n$, decay width $\Gamma_n$ and 4-momentum $k$ in the unitary gauge is:
\be
i \Delta^G_{\mu \nu \alpha \beta}(k) = \frac{i P_{\mu \nu \alpha \beta}(k,m_n)}{k^2-m^2_n+i m_n \Gamma_n} \, , 
\ee
where $P_{\mu \nu \alpha \beta}$ is the sum of the polarization tensors $\epsilon^s_{\mu \nu} (k)$ (being $s$ the spin):
\begin{eqnarray}
P_{\mu \nu \alpha \beta}(k,m_g) & = & \sum_s \epsilon^s_{\mu \nu}(k)\epsilon^s_{\alpha \beta}(k) \nonumber \\
&& \nonumber \\
&=& \frac{1}{2}(G_{\mu \alpha}G_{\nu \beta} + G_{\nu \alpha}G_{\mu \beta} - \frac{2}{3}G_{\mu \nu}G_{\alpha \beta} )
\end{eqnarray}
and
\be
G_{\mu \nu} \equiv \eta_{\mu \nu} - \frac{k_\mu k_\nu}{m^2_n} \, .
\ee
The tensor $P_{\mu \nu \alpha \beta}$ must satisfy several conditions for an on-shell graviton $G_{\mu\nu}$, 
in order to reduce the number of degrees-of-freedom to the physical ones:
\be
\eta^{\alpha\beta}P_{\mu \nu \alpha \beta}(k,m_g)=\eta^{\nu\mu}P_{\mu \nu \alpha \beta}(k,m_n)=0 \, ,
\ee
\be
k^{\alpha}P_{\mu \nu \alpha \beta}(k,m_g)=k^{\beta}P_{\mu \nu \alpha \beta}(k,m_g)=k^{\mu}P_{\mu \nu \alpha \beta}(k,m_g)=k^{\nu}P_{\mu \nu \alpha \beta}(k,m_g)=0 \, .
\ee

\section{Feynman rules}
\label{app:feynman}

We summarize  in this Appendix the different Feynman rules corresponding to the couplings of scalar DM particles and of SM particles  with KK-gravitons and radions. 

\subsection{Graviton Feynman rules}
\label{app:gravFR}

The vertex that involves one KK-graviton (with $n \neq 0$) and two scalars of mass $m_S$ is given by:
\be
\qquad
\raisebox{-15mm}{\includegraphics[keepaspectratio = true, scale = 1] {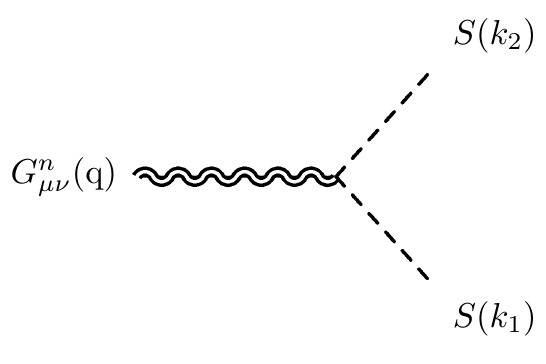}}
\begin {aligned}
=-\frac{i}{\Lambda} \left ( m^2_S \eta_{\mu \nu} - C_{\mu \nu \rho \sigma} k_1^{\rho} k_2^{\sigma} \right ) \, ,
\end {aligned}
\ee
where
\be
C_{\mu \nu \alpha \beta} \equiv \eta_{\mu \alpha} \eta_{\nu \beta} + \eta_{\nu \alpha} \eta_{\mu \beta} - \eta_{\mu \nu} \eta_{\alpha \beta} \, .
\ee
This expression can be used for the coupling of both scalar DM and the SM Higgs boson to KK-gravitons.

The Feynman rule corresponding to the interaction of two SM Dirac fermions of mass $m_\psi$ with one KK-graviton is given by: 
\be
\qquad
\raisebox{-15mm}{\includegraphics[keepaspectratio = true, scale = 1] {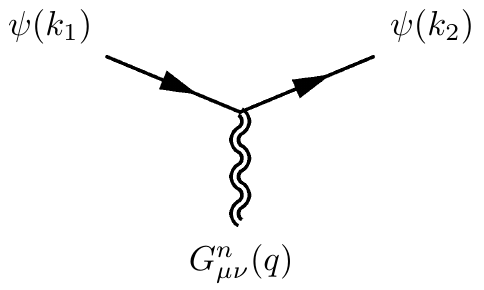}}
\begin {aligned}
=& - \frac{i}{4\Lambda}
\left [ \gamma_{\mu} \left ( k_{2 \nu}+k_{1 \nu} \right ) + \gamma_{\nu} \left ( k_{2 \mu}+k_{1 \mu} \right ) \right. \\
&\left. - 2 \eta_{\mu \nu}\left ( \slashed{k_2}+\slashed{k_1}-2m_{\psi} \right )\right ] \, ,
\end {aligned}
\ee
whereas
\be
\qquad
\raisebox{-15mm}{\includegraphics[keepaspectratio = true, scale = 1] {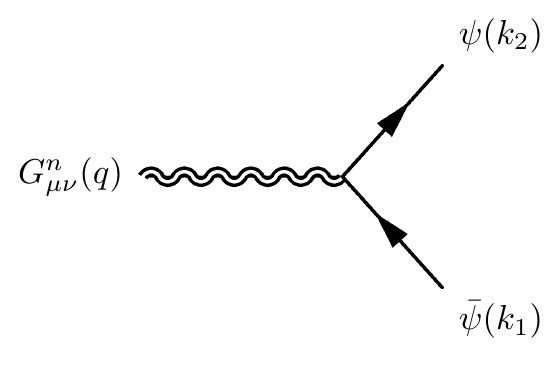}}
\begin {aligned}
=& - \frac{i}{4\Lambda}\left [ \gamma_{\mu} \left (k_{2 \nu} - k_{1 \nu} \right ) 
+ \gamma_{\nu} \left (k_{2 \mu} - k_{1 \mu} \right ) \right. \\
&\left. - 2 \eta_{\mu \nu}\left ( \slashed{k_2} -\slashed{k_1} -2m_{\psi} \right )\right ] \, .
\end {aligned}
\ee

The interaction between two SM gauge bosons of mass $m_A$ and one KK-graviton is given by:
\be
\qquad
\raisebox{-15mm}{\includegraphics[keepaspectratio = true, scale = 1] {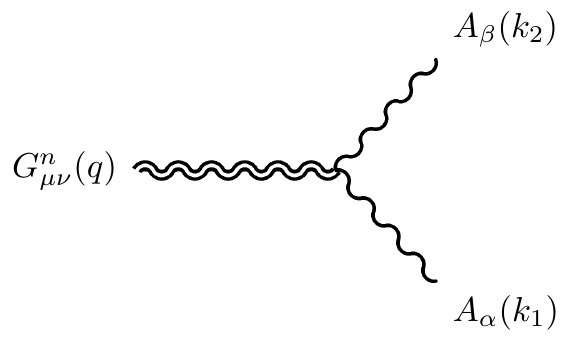}}
\begin {aligned}
=-\frac{i}{\Lambda} \left ( m^2_A C_{\mu \nu \alpha \beta} + W_{\mu \nu \alpha \beta} \right ) \, ,
\end {aligned}
\ee
where
\be
W_{\mu \nu \alpha \beta} \equiv B_{\mu \nu \alpha \beta} + B_{\nu \mu \alpha \beta}
\ee
and
\be
B_{\mu \nu \alpha \beta} \equiv \eta_{\alpha \beta}k_{1 \mu}k_{2 \nu} + \eta_{\mu \nu}(k_1 \cdot k_2 \, \eta_{\alpha \beta} - k_{1 \beta} k_{2 \nu}) - \eta_{\mu \beta} k_{1 \nu} k_{2 \alpha} + \frac{1}{2}\eta_{\mu \nu}(k_{1 \beta} k_{2 \alpha} - k_1 \cdot k_2 \, \eta_{\alpha \beta}) \, .
\ee

Eventually, the interaction between two scalar DM particles and two KK-gravitons (coming from a second order expansion of the metric $g_{\mu\nu}$ about the Minkowski metric $\eta_{\mu\nu}$) is given by:
\be
\qquad
\raisebox{-15mm}{\includegraphics[keepaspectratio = true, scale = 1] {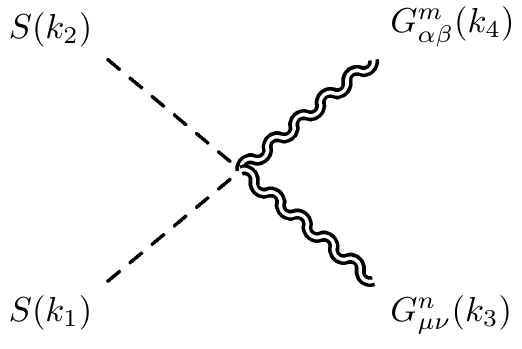}}
\begin {aligned}
=& -\frac{i}{\Lambda^2} \eta_{\nu \beta} \left ( m^2_S \eta_{\mu \alpha} - C_{\mu \alpha \rho \sigma} k_1^{\rho} k_2^{\sigma} \right ) \, .
\end {aligned}
\ee

The Feynman rules for the $n=0$ KK-graviton can be obtained by the previous ones by replacing $\Lambda$ with $M_{\rm P}$. 
We do not give here the triple KK-graviton vertex, as it is irrelevant for the phenomenological applications of this paper. 
The same occurs for the vertices between one KK-graviton and two radions and two KK-gravitons and one radion.

\subsection{Radion Feynman rules}
\label{app:radFR}

The radion field $r$ couples with both the SM and the DM particles with the trace of the energy-momentum tensor, $T = g^{\mu\nu} T_{\mu\nu}$. 
The only exception are photons and gluons that, being massless, do not contribute to $T$ at tree-level. 
However,  effective couplings of these fields to the radion are generated  through quarks and $W$ loops, and the trace anomaly.

The interaction between one radion and two scalar fields (either the DM or the SM Higgs boson) is given by:
\be
\qquad
\raisebox{-15mm}{\includegraphics[keepaspectratio = true, scale = 1] {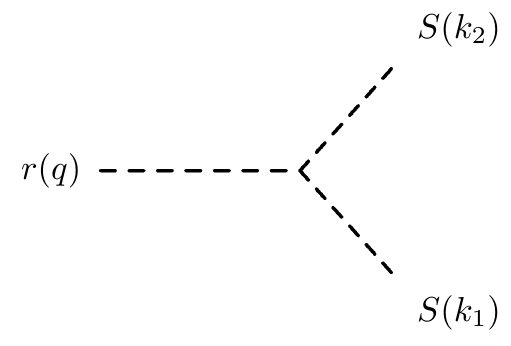}}
\begin {aligned}
=-\frac{2i}{\Lambda\sqrt{6}} \left ( 2m^2_S + k_{1\mu} k_2^\mu \right ) \, .
\end {aligned}
\ee

The vertex that involves the radion and two SM Dirac fermions takes the form:
\be
\qquad
\raisebox{-15mm}{\includegraphics[keepaspectratio = true, scale = 1] {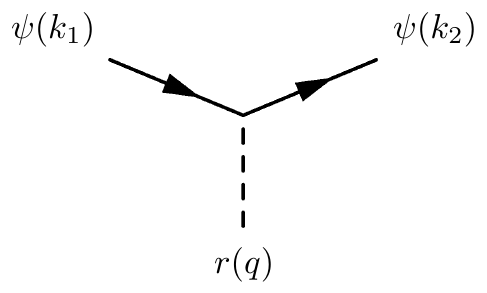}}
\begin {aligned}
=- \frac{i}{2\Lambda\sqrt{6}} \left [8m_{\psi} - 3 \left ( \slashed{k_2}+\slashed{k_1} \right ) \right]  
\end {aligned}
\ee
and, as in the case of the graviton-fermion-fermion vertex, we have:
\be
\qquad
\raisebox{-15mm}{\includegraphics[keepaspectratio = true, scale = 1] {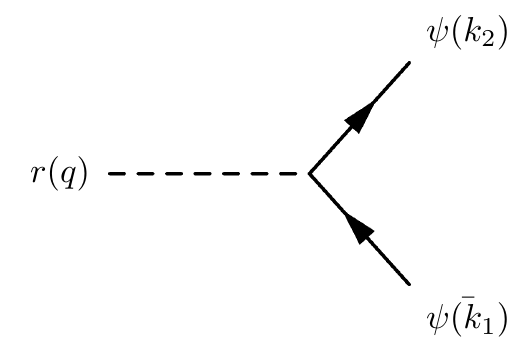}}
\begin {aligned}
=- \frac{i}{2\Lambda\sqrt{6}} \left [ 8m_{\psi} - 3 \left ( \slashed{k_2}-\slashed{k_1} \right ) \right] \, .
\end {aligned}
\ee

The interaction between two massive SM gauge bosons and one radion is given by: 
\be
\qquad
\raisebox{-15mm}{\includegraphics[keepaspectratio = true, scale = 1] {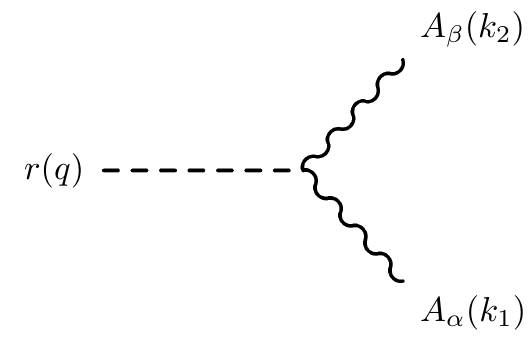}}
\begin {aligned}
=\frac{2i}{\Lambda\sqrt{6}} m^2_A \eta_{\alpha \beta} \, .
\end {aligned}
\ee

The Feynman rule corresponding to the interaction between two massless SM gauge bosons and one radion is: 
\be
\qquad
\raisebox{-15mm}{\includegraphics[keepaspectratio = true, scale = 1] {radion_vector_vector.pdf}}
\begin {aligned}
\label{eq:radiontomasslessvertex}
=\frac{4i\alpha_{i}C_{i}}{8\pi\Lambda\sqrt{6}} \left [ \eta_{\mu\nu}(k_1 \cdot k_2) - k_{1\nu}k_{2\mu}  \right ] \, ,
\end {aligned}
\ee
where $\alpha_{i}=\alpha_{EM}, \alpha_s$ for the case of the photons or gluons, respectively, and
\be
\left \{
\begin{array}{lll}
C_3 &=& b_{IR}^{(3)} - b_{UV}^{(3)} + \frac{1}{2}\sum_q F_{1/2}(x_q) \, , \\
&&\\
C_{EM} &=& b_{IR}^{(EM)} - b_{UV}^{(EM)} + F_1(x_W)  - \sum_q N_cQ_{q}^2F_{1/2}(x_q) \, ,
\end{array}
\right .
\ee
with $x_q = 4m_q/m_r$ and $x_W = 4m_w/m_r$. The explicit form of $F_{1/2}$  and the values of the one-loop $\beta$-function coefficients $b$ are given by \cite{Blum:2014jca}:
\be
\left \{
\begin{array}{lll}
F_{1/2}(x) = 2x[1 + (1-x)f(x)] , \\
&&\\
F_{1}(x) = 2 + 3x + 3x(2-x)f(x) , 
\end{array}
\right .
\ee
\be
f(x) = \left \{
\begin{array}{lll}
[\arcsin(1/\sqrt{x})]^2 \hphantom{} \hphantom{} \hphantom{} \hphantom{} \hphantom{} &x>1 , \\
&&\\
-\frac{1}{4}\left[\log\left( \frac{1 + \sqrt{x-1}}{1 - \sqrt{x-1}} \right) - i\pi \right]^2 &x<1 ,
\end{array}
\right .
\ee
while $b_{IR}^{(EM)} - b_{UV}^{(EM)} = 11/3$ and $b_{IR}^{(3)} - b_{UV}^{(3)} = -11 + 2n/3$, where $n$ is the number of quarks whose mass is smaller than $m_r/2$.

Eventually, the interaction Lagrangian between the DM and the radion up to second order is given by
\footnote{In the second order interaction terms for the radion, based on \cite{Goldberger:1999un},  we have found some numerical factors that differ from 
Refs. \cite{Csaki:1999mp, Lee:2013bua}, however such difference will not modify  our main  results, since the dominant DM annihilation channel in most of the allowed region is into KK-gravitons.}  :
\be
\mathcal{L} = \frac{1}{\Lambda \sqrt{6}} r T^{\rm DM} - \frac{1}{12\Lambda^2}r^2 (\partial_\mu S) (\partial\mu S) + \frac{1}{2\Lambda^2} r^2 S^2 \, ,
\ee
being $T^{\rm DM}$ the trace of the energy-momentum tensor of the DM eq.~(\ref{Tensor_SM_DM}). As in the case of the interactions with gravitons, exists a 4-legs interaction term:
\be
\label{eq:4pointsSSrr}
\qquad
\raisebox{-15mm}{\includegraphics[keepaspectratio = true, scale = 1] {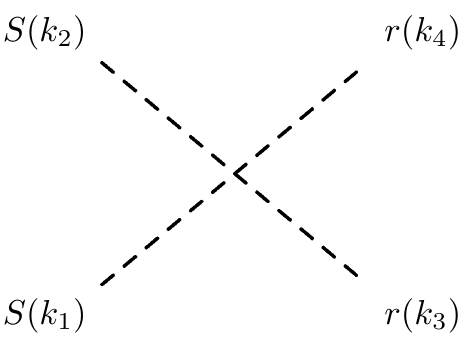}}
\begin {aligned}
=- \frac{i}{3\Lambda^2} \left( 6m_{S}^2 + k_{1\mu} k_2^\mu \right)  \, .
\end {aligned}
\ee

\section{Decay widths}
\label{app:decay}

In this appendix we compute the decay widths of KK-gravitons and of the radion, using the Feynman rules given in App.\ref{app:feynman}. 

\subsection{KK-graviton decay widths}
\label{app:gravdecay}

The KK-graviton can decay into scalar particles (including the Higgs boson, the DM particle, if the mass of the considered KK-graviton is sufficiently large, 
and the radion), SM fermions, SM gauge bosons and lighter KK-gravitons.

Decay widths of KK-gravitons into SM particles, $\Gamma (G_n \to {\rm SM} \, {\rm SM})$, are all proportional to $1/\Lambda^2$. 
In particular, the decay width into SM Higgs bosons  is given by:
\be
\Gamma(G_n \rightarrow hh) = \frac{m_n^3}{960 \pi\Lambda^2}\left(1-\frac{4m_h^2}{m_n^2}\right)^{5/2} \, ,
\ee
where $m_n$ is the mass of the $n$-th KK-graviton (in the main text, this was called $m_{G_n}$, but we prefer here a shorter notation to increase 
readability of the formul\ae).
If $m_n > 2 m_S$, the $n$-th KK-graviton can decay into two DM particles: 
\be
\Gamma(G \rightarrow SS) = \frac{m_n^3}{960 \pi\Lambda^2}\left(1-\frac{4m_S^2}{m_n^2}\right)^{5/2} \, .
\ee

The decay width of the $n$-th KK-graviton into SM Dirac fermions is given by: 
\be
\Gamma(G_n \rightarrow \bar{\psi} \psi) = \frac{m_n^3}{160 \pi\Lambda^2}\left( 1-\frac{4 m_{\psi}^2}{m_n^2} \right)^{3/2}\left( 1+\frac{8 m_{\psi}^2}{3 m_n^2} \right).
\ee

The decay width of the $n$-th KK-graviton into two SM massive gauge bosons reads: 
\be
\left \{
\begin{array}{lll}
\Gamma(G_n \rightarrow W^+ W^-) &=& 
\frac{m_n^3}{480 \pi\Lambda^2} \left( 1-\frac{4 m_W^2}{m_n^2}\right)^{1/2}\left( 13+\frac{56 m_W^2}{m_n^2} + \frac{48 m_W^4}{m_n^4} \right) \, , \\
&& \\
\Gamma(G_n \rightarrow ZZ) &=& \frac{m_n^3}{960 \pi\Lambda^2} \left( 1-\frac{4 m_Z^2}{m_n^2}\right)^{1/2}\left( 13+\frac{56 m_Z^2}{m_n^2} + \frac{48 m_Z^4}{m_n^4} \right)
\, ,
\end{array}
\right .
\ee
whereas the decay width into massless gauge bosons is:
\be
\left \{
\begin{array}{lll}
\Gamma(G_n \rightarrow \gamma \gamma) &=& \frac{m_n^3}{80 \pi\Lambda^2} \, ,\\
& & \\
\Gamma(G_n \rightarrow gg) &=& \frac{m_n^3}{10 \pi\Lambda^2} \, .
\end{array}
\right .
\ee

On the other hand, the decay widths of KK-gravitons with KK-number $n$ into lighter KK-gravitons are proportional to $1/\Lambda^6$, as the triple
graviton vertex comes from the third order expansion of the metric about the Minskowski spacetime. For this reason, we have not considered
these decays when computing the total KK-graviton decay widths. The same happens for the radion: the coupling of the radion with the gravitons
arises from the mixing of the radion with the graviscalar $h_{55}$, that eventually couples with KK-gravitons again with a triple graviton vertex, 
proportional to $1/\Lambda^3$. Also in this case the decay width $\Gamma (G_n \to r \, r)$ is proportional to $1/\Lambda^6$ and, therefore, negligible. 

\subsection{Radion decay widths}
\label{app:radion_decay}

The decay width of the radion into scalar particles, either the SM Higgs boson or the DM particle if the radion is sufficiently heavy, is given by:
\be
\Gamma(r \rightarrow hh, SS) = \frac{m_r^3}{192 \pi\Lambda^2}\left(1-\frac{4m_X^2}{m_r^2}\right)^{1/2}\left(1+\frac{2m_X^2}{m_r^2}\right)^{2} \, ,
\ee
where $m_X = m_h, m_S$ depending on the considered channel.

The radion decay width into SM Dirac fermions is given by:
\be
\Gamma(r \rightarrow \bar{\psi} \psi) = \frac{m_r m^2_{\psi}}{48 \pi\Lambda^2}\left( 1-\frac{4 m_{\psi}^2}{m_r^2} \right)^{3/2} \, .
\ee

The radion decay width into SM massive gauge bosons reads: 
\be
\left \{
\begin{array}{lll}
\Gamma(r \rightarrow W^+W^-) &=& \frac{m_r^3}{96 \pi\Lambda^2}\left(1-\frac{4m_W^2}{m_r^2}\right)^{1/2}\left(12 - \frac{4m_W^2}{m_r^2} + \frac{m_W^4}{m_r^4}\right) \, , \\
&&\\
\Gamma(r \rightarrow ZZ) &=& \frac{m_r^3}{192 \pi\Lambda^2}\left(1-\frac{4m_Z^2}{m_r^2}\right)^{1/2}\left(12 - \frac{4m_Z^2}{m_r^2} + \frac{m_Z^4}{m_r^4}\right) \, ,
\end{array}
\right .
\ee
whereas the decay width into SM massless gauge bosons is: 
\be
\left \{
\begin{array}{lll}
\Gamma(r \rightarrow \gamma \gamma) &=& \frac{\alpha_{EM}C_{EM}m_r^3}{7680 \pi\Lambda^2} \, , \\
&&\\
\Gamma(r \rightarrow g g) &=& \frac{\alpha_{3}C_{3}m_r^3}{960 \pi\Lambda^2} \, .
\end{array}
\right .
\ee

\section{Annihilation DM Cross section}
\label{app:annihil}

Since in the freeze-out scenario, DM annihilation occurs at small relative velocity of the two DM particles, it is useful to  approximate the Mandelstam variable $s$ as: 
\be
s \approx m_{dm}^2(4 + v_{rel}^2) \, .
\ee
Within this approximation, the different scalar products for processes in which two DM particles $S$'s annihilate into two SM particles $X$'s, with incoming and outcoming momenta $S(k_1) \, S(k_2) \to X(k_3) \, X(k_4)$, become:
\be
\left \{
\begin{array}{lll}
k_1 \cdot k_4 = k_2 \cdot k_3 \approx 
m_{S}^2 + \frac{1}{2}m_{S}^2\sqrt{1-\frac{m_{X}^2}{m_{S}^2}}\, \cos{\theta} \, v_{rel} + \frac{1}{4} m_{S}^2 \, v_{rel}^2 \, , \\
&&\\
k_1 \cdot k_3 = k_2 \cdot k_4 \approx 
m_{S}^2 - \frac{1}{2}m_{S}^2\sqrt{1-\frac{m_{X}^2}{m_{S}^2}}\, \cos{\theta} \, v_{rel} + \frac{1}{4} m_{S}^2 \, v_{rel}^2 \, ,
\end{array}
\right .
\ee
where
\be
\left \{
\begin{array}{lll}
k_1 \cdot k_1 &=& k_2 \cdot k_2 = m_S^2 \, , \\
&& \\
k_3 \cdot k_3 &=& k_4 \cdot k_4 = m_X^2 \, .
\end{array}
\right .
\ee
We will always write the annihilation cross-sections at leading order  in this  velocity expansion.

\subsection{Annihilation through and into Gravitons}
\label{app:scalarannihil}

The annihilation of DM particles  into SM particles through virtual KK-graviton exchange occurs in d-wave. 
In the following expressions, $S_{KK}$ stands for the sum over all KK states:
\be
S_{KK} = \frac{1}{\Lambda^2}\sum_{n=1}^{\infty} \frac{1}{s-m_n^2+i m_n \Gamma_n} \, ,
\ee
where $m_n$ is the mass of the $n$-th KK-graviton.
 
\noindent The annihilation cross-section into two SM Higgs bosons reads:
\be
\sigma_g(S \, S \rightarrow h \, h) \approx v_{rel}^3 \cdot |S_{KK}|^2 \frac{m_S^6}{720 \pi} \left(1-\frac{m_h^2}{m_S^2}\right)^{5/2} \, .
\ee
The annihilation cross-section into two SM massive gauge bosons is given by:
\be
\left \{
\begin{array}{lll}
\sigma_g(S \, S \rightarrow W^+ \, W^-) &\approx& v_{rel}^3 \cdot |S_{KK}|^2 \frac{m_S^6}{360 \pi}\left( 1-\frac{m_w^2}{m_S^2} \right)^{1/2} \left( 13 + \frac{14 m_w^2}{m_S^2} + \frac{3 m_w^4}{m_S^4} \right) \, , \\
&&\\
\sigma_g(S \, S \rightarrow Z \, Z) &\approx & v_{rel}^3 \cdot |S_{KK}|^2 \frac{m_S^6}{720 \pi}\left( 1-\frac{m_w^2}{m_S^2} \right)^{1/2} \left( 13 + \frac{14 m_Z^2}{m_S^2} + \frac{3 m_Z^4}{m_S^4} \right)  \, ,
\end{array}
\right .
\ee
whereas for two massless gauge bosons we have:
\be
\left \{
\begin{array}{lll}
\sigma_g(S \, S \rightarrow \gamma \, \gamma) &\approx& v_{rel}^3 \cdot |S_{KK}|^2 \frac{m_S^6}{60 \pi} \, , \\
&&\\
\sigma_g(S \, S \rightarrow g \, g) &\approx& v_{rel}^3 \cdot |S_{KK}|^2 \frac{2 m_S^6}{15 \pi} \, .
\end{array}
\right .
\ee
Eventually, the annihilation cross-section into two SM fermions is:
\be
\sigma_g(S \, S \rightarrow \bar{\psi} \, \psi) \approx v_{rel}^3 \cdot |S_{KK}|^2 \frac{m_s^6}{360 \pi}\left( 1 - \frac{m_{\psi}^2}{m_s^2} \right)^{3/2}\left( 3 + \frac{2m_{\psi}^2}{m_s^2} \right) \, .
\ee

As it was shown in Ref.~\cite{Lee:2013bua}, for DM particle masses larger than the mass of a given KK-graviton mode
DM particles may annihilate into two KK-gravitons. In the small velocity approximation, the corresponding cross-section is:
\bea
\sigma_g ( S \, S \rightarrow G_n \, G_m) \approx v_{rel}^{-1} \, \left ( \frac{A + B + C/4}{9216 \pi } \right ) \, 
\left ( \frac{1}{\Lambda ^4 \, m_{\rm S}^3 \, m_{\rm n}^4 \, m_{\rm m}^4} \right ) \, \sqrt{\frac{\left(4 m_{\text{S}}^2+m_{\text{n}}^2-m_{\text{m}}^2\right){}^2}{16
   m_{\text{S}}^2}-m_{\text{n}}^2} \, , \nonumber \\ 
   \hphantom{} 
\label{eq:graviton_real_scalar_dm_cruzados}
\eea
where the three contributions to the cross-section come from the square of the $t$- and $u$-channels 
amplitudes in diagrams (a) and (b) of Fig.~\ref{fig:diagramas} (A), the square of the 4-points amplitude in 
diagram (c) of the same Figure (C) and from the interference between the two classes of diagrams (B), respectively:
\begin{equation}
\left \{
\begin{array}{lll}
A &=& \frac{\left [
-2 \, m_{\rm m}^2 \, \left ( 4 \, m_{\rm S}^2 + m_{\rm n}^2 \right )
+ \left(m_{\rm n}^2 - 4 \, m_{\rm S}^2 \right)^2 
+ m_{\rm m}^4 \right ]^4}{2 \left(4 \, m_{\rm S}^2 - m_{\rm n}^2 - m_{\rm m}^2 \right)^2} \, , \\
  && \\
B &=& \frac{ \left [
-8 \, m_{\rm S}^2 \, \left(m_{\rm n}^2 + m_{\rm m}^2 \right)+16 \, m_{\rm S}^4 + 
\left(m_{\rm n}^2 - m_{\rm m}^2 \right)^2 \right ]^2}{4 \, m_{\rm S}^2 - m_{\rm n}^2 - m_{\rm m}^2} \,
   \left[16 \, m_{\rm S}^4
   \left(m_{\rm n}^2 + m_{\rm m}^2 \right) \right.  \\
   &-& \left. 8 \, m_{\rm S}^2 \, \left( - m_{\rm n}^2 \,
   m_{\rm m}^2 + m_{\rm n}^4 + m_{\rm m}^4 \right) +\left( m_{\rm n}^2 - m_{\rm m}^2\right)^2 \,
   \left(m_{\rm n}^2 + m_{\rm m}^2 \right) \right]  \, , \\
  && \\
C &=& 256 \, m_{\rm S}^8 \, \left(13 \, m_{\rm n}^2 \,  m_{\rm m}^2 + 2 \, m_{\rm n}^4 + 2 \, m_{\rm m}^4 \right) 
- 512 \, m_{\rm S}^6 \, \left(m_{\rm n}^6 + m_{\rm m}^6 \right)  \\   
&&\\
   &+& 32 \, m_{\rm S}^4 \left (-17 \, m_{\rm n}^6 \, m_{\rm m}^2
   + 98 \,  m_{\rm n}^4 \, m_{\rm m}^4 - 17 \, m_{\rm n}^2 \, m_{\rm m}^6 + 6 \, m_{\rm n}^8 + 6 \, m_{\rm m}^8 \right) \\
&&\\
   &-& 32 \, m_{\rm S}^2  \left(m_{\rm n}^2 - m_{\rm m}^2 \right)^2 \,
   \left(m_{\rm n}^6 + m_{\rm m}^6 \right) 
   + \left(m_{\rm n}^2 - m_{\rm m}^2 \right)^4 \, \left(13 \, m_{\rm n}^2 \, 
   m_{\rm m}^2 + 2 \, m_{\rm n}^4 + 2 \, m_{\rm m}^4 \right) \, . \\
  &&
\end{array}
\right .
\end{equation}

When the two KK-gravitons have the same KK-number, $m = n$,
 eq.~(\ref{eq:graviton_real_scalar_dm_cruzados}) simplifies:
\begin{eqnarray}
\sigma_g ( S \, S \rightarrow G_n \, G_n) & \approx &v_{rel}^{-1} \frac{m_S^2}{576 \, \pi  \, \Lambda^4} 
\frac{(1-r)^{1/2}}{r^4(2-r)^2}  \, \left ( 256 - 768 \, r + 968 \, r^2 - 520 \, r^3 \right . \nonumber \\
&+& \left .  142 \, r^4 - 52 \, r^5 + 19 \, r^6 \right ) \, ,
\label{graviton_real_scalar_dm}
\end{eqnarray}
where $r \equiv (m_{\rm n }/m_S)^2$.

\subsection{Annihilation through and into Radions}
\label{app:scalarannihilrad}

When the distance between the two branes is stabilized using the Goldberger-Wise mechanism, 
the DM particles can annihilate into SM particles also through virtual radion exchange. 
The processes involving the radion occur in S-wave and can be more efficient than the exchange 
of a tower of virtual KK-gravitons, which is in d-wave.

The DM annihilation cross-section into the SM Higgs boson is: 
\be
\sigma_r (S \, S \rightarrow h \, h) \approx v_{rel}^{-1} \frac{m_S^6}{16 \, \pi \, \Lambda^4} 
\, \frac{1}{(s - m_r^2)^2 + m_r^2 \, \Gamma_r^2} \, \left(1 - \frac{m_h^2}{m_S^2} \right)^{1/2} \, 
\left(2 + \frac{m_h^2}{m_S^2}\right)^2 \, ,
\ee
where $m_r$ is the mass of the radion.

The cross-section for DM annihilation into SM massive gauge bosons reads: 
\be
\left \{
\begin{array}{lll}
\sigma_r (S \, S \rightarrow W^+ \, W^-) &\approx & v_{rel}^{-1} \, \frac{m_S^6}{8 \, \pi \, \Lambda^4} \, 
\frac{1}{(s-m_r^2)^2 + m_r^2 \, \Gamma_r^2}  
 \,  \left( 1 - \frac{m_w^2}{m_S^2} \right)^{1/2} \, \left( 4 - \frac{4 m_w^2}{m_S^2} + \frac{3 \, m_w^4}{m_S^4} \right ) \, , \\
&& \\
\sigma_r (S \, S \rightarrow Z \, Z) & \approx & v_{rel}^{-1} \, \frac{m_S^6}{16 \, \pi \, \Lambda^4} \, 
\frac{1}{(s - m_r^2)^2 + m_r^2\, \Gamma_r^2} 
\,  \left( 1 - \frac{m_w^2}{m_S^2} \right)^{1/2} \, \left( 4 - \frac{4 m_Z^2}{m_S^2} + \frac{3 \, m_Z^4}{m_S^4} \right) \, . \\
&&
\end{array}
\right .
\ee
The DM annihilation into photons and gluons is proportional to the vertex in eq.~(\ref{eq:radiontomasslessvertex}). 
The corresponding expressions for the cross-sections are:
\be
\left \{
\begin{array}{lll}
\sigma_r (S \, S \rightarrow \gamma \, \gamma) &\approx & v_{rel}^{-1} \, 
\frac{m_S^6 \, \alpha_{EM} \, C_{EM}}{32 \, \pi^3 \, \Lambda^4} \, \frac{1}{(s - m_r^2)^2 + m_r^2 \, \Gamma_r^2} \, , \\
&& \\
\sigma_r (S \, S \rightarrow g \, g) & \approx & v_{rel}^{-1} \, 
\frac{m_S^6 \, \alpha_{3} \, C_{3}}{4 \, \pi^3 \, \Lambda^4} \, \frac{1}{(s - m_r^2)^2 + m_r^2\, \Gamma_r^2} \, .
\end{array}
\right .
\ee
Eventually, the DM annihilation cross-section into SM fermions is given by:
\be
\sigma_r (S \, S \rightarrow \bar{\psi} \, \psi) \approx v_{rel}^{-1} \, \frac{m_s^4 \, m_{\psi}^2}{4 \, \pi \, \Lambda^4} \, 
\frac{1}{(s - m_r^2)^2 + m_r^2\, \Gamma_r^2} \, \left( 1 - \frac{m_{\psi}^2}{m_s^2} \right)^{3/2}.
\ee

As in the case of the graviton, if the mass of the DM is larger than the mass of the radion, then the DM particles
can annihilate into two on-shell radions:
\be
\sigma_r (S \, S \rightarrow r \, r) \approx v_{rel}^{-1} \, 
\frac{\, m_S^5 \, \sqrt{m_S^2 - m_r^2} }{576 \, \pi \, \Lambda^4 \, \left( m_r^2 - 2 \, m_S^2 \right)^2}\left(2 + 7 \, \frac{m_r^2}{m_S^2} \right)^2 \, ,
\ee
where we have considered both the $u$- and $t$-channels amplitudes and the contribution coming from the 4-legs vertex in eq.~(\ref{eq:4pointsSSrr}).

\bibliography{bibliografia} 

\end{document}